\documentclass[twoside,english]{elsarticle}
\usepackage[LGR,T1]{fontenc}
\usepackage[latin9]{inputenc}
\usepackage{babel}
\usepackage{array}
\usepackage{float}
\usepackage{textcomp}
\usepackage{graphicx}
\usepackage{setspace}
\usepackage{esint}
\usepackage[unicode=true]{hyperref}

\makeatletter

\def\ps@pprintTitle{%
 \let\@oddhead\@empty
 \let\@evenhead\@empty
 \def\@oddfoot{{\footnotesize\itshape
 Pattern Recognition (accepted manuscript)\hfill\today}}%
 \let\@evenfoot\@oddfoot}

\DeclareRobustCommand{\greektext}{%
  \fontencoding{LGR}\selectfont\def\encodingdefault{LGR}}
\DeclareRobustCommand{\textgreek}[1]{\leavevmode{\greektext #1}}
\ProvideTextCommand{\~}{LGR}[1]{\char126#1}

\providecommand{\tabularnewline}{\\}
\floatstyle{ruled}
\newfloat{algorithm}{tbp}{loa}
\providecommand{\algorithmname}{Algorithm}
\floatname{algorithm}{\protect\algorithmname}

\journal{Pattern Recognition}

\@ifundefined{showcaptionsetup}{}{%
 \PassOptionsToPackage{caption=false}{subfig}}
\usepackage{subfig}
\makeatother

\begin{document}
\def\appendixname{Appendix } 

\begin{frontmatter}{}

\title{The Partially Observable Hidden Markov Model \\
and its Application to Keystroke Dynamics}

\author[arl]{John V.~Monaco\corref{cor1}}

\ead{john.v.monaco2.civ@mail.mil}

\ead[url]{http://www.vmonaco.com}

\author[pace]{Charles C.~Tappert}

\cortext[cor1]{Corresponding author}

\address[arl]{U.S.~Army Research Laboratory, Aberdeen Proving Ground, MD 21005,
USA}

\address[pace]{Pace University, Pleasantville, NY 10570, USA}
\begin{abstract}
The partially observable hidden Markov model is an extension of the
hidden Markov Model in which the hidden state is conditioned on an
independent Markov chain. This structure is motivated by the presence
of discrete metadata, such as an event type, that may partially reveal
the hidden state but itself emanates from a separate process. Such
a scenario is encountered in keystroke dynamics whereby a user's typing
behavior is dependent on the text that is typed. Under the assumption
that the user can be in either an active or passive state of typing,
the keyboard key names are event types that partially reveal the hidden
state due to the presence of relatively longer time intervals between
words and sentences than between letters of a word. Using five public
datasets, the proposed model is shown to consistently outperform other
anomaly detectors, including the standard HMM, in biometric identification
and verification tasks and is generally preferred over the HMM in
a Monte Carlo goodness of fit test.
\end{abstract}

\end{frontmatter}{}

\section{Introduction\label{sec:intro}}

The hidden Markov model (HMM), which dates back over 50 years \citep{baum1966statistical},
has seen numerous applications in the recognition of human behavior,
such as speech \citep{rabiner1989tutorial}, gesture \citep{yamato1992recognizing},
and handwriting \citep{hu1996hmm}. Recent successes have leveraged
the expressive power of connectionist models by combining the HMM
with feed-forward deep neural networks, which are used to estimate
emission probabilities \citep{espana2011improving,hinton2012deep,wu2016deep}.
Despite the increasing interest in sequential deep learning techniques,
e.g., recurrent neural networks, HMMs remain tried-and-true for time
series analyses. The popularity and endurance of the HMM can be at
least partially attributed to the tractability of core problems (parameter
estimation and likelihood calculation), ability to be combined with
other methods, and the level of insight it provides to the data.

At least part its success can also be attributed to its flexibility,
with many HMM variants having been developed for specific applications.
This usually involves introducing a dependence, whether it be on time
\citep{yu2010hidden}, previous observations \citep{LI2005977}, or
a semantic context \citep{bianne2011dynamic}. The motivation for
doing so is often to better reflect the structure of the underlying
problem. Although many of these variations have increased complexity
and number of parameters over the standard HMM, their estimation remains
tractable.

In this work, we introduce the partially observable hidden Markov
model (POHMM), an extension of the HMM intended for keystroke dynamics.
We are interested in modeling the temporal behavior of a user typing
on a keyboard, and note that certain keyboard keys are thought to
influence typing speed. Non-letter keys, such as punctuation and the
Space key, indicate a greater probability of being in a \emph{passive}
state of typing, as opposed to an \emph{active} state, since the typist
often pauses between words and sentences as opposed to between letters
in a word \citep{salthouse1986perceptual}. The POHMM reflects this
scenario by introducing a dependency on the key names which are observed
alongside the time intervals, and in this way, the keys provide a
context for the time intervals.

The idea of introducing a context upon which some behavior depends
is not new. Often, an observation depends not only on a latent variable
but on the observations that preceded it. For example, the neighboring
elements in a protein secondary structure can provide context for
the element under consideration, which is thought to depend on both
the previous element and a hidden state \citep{LI2005977}; nearby
phonemes can aid in the recognition of phonemes \citep{lee1989speaker};
and the recognition of human activities can be achieved with greater
accuracy by considering both a spatial context (e.g., where the activity
occurred) and temporal context (e.g., the duration of the activity)
\citep{CHUNG20081572}.

Handwriting recognition has generally seen increased performance with
models that consider the surrounding context of a handwritten character.
The rationale for such an approach is that a character may be written
with different style or strokes depending on its neighboring characters
in the sequence. Under this assumption, the neighboring pixels or
feature vectors of neighboring characters can provide additional context
for the character under consideration. Alternatively, a separate model
can be trained for each context in which the character appears, e.g.,
``t'' followed by ``e'' versus ``t'' followed by ``h'' \citep{bianne2011dynamic}.
This same principle motivates the development of the POHMM, with the
difference being that the context is provided not by the observations
themselves, but by a separate sequence.

We apply the POHMM to address the problems of user identification,
verification, and continuous verification, leveraging keystroke dynamics
as a behavioral biometric. Each of these problems requires estimating
the POHMM parameters for each individual user. Identification is performed
with the maximum a posteriori (MAP) approach, choosing the model with
maximum a posterior probability; verification, a binary classification
problem, is achieved by using the model log-likelihood as a biometric
score; and continuous verification is achieved by accumulating the
scores within a sliding window over the sequence. Evaluated on five
public datasets, the proposed model is shown to consistently outperform
other leading anomaly detectors, including the standard HMM, in biometric
identification and verification tasks and is generally preferred over
the HMM in a Monte Carlo goodness of fit test.

All of the core HMM problems remain tractable for the POHMM, including
parameter estimation, hidden state prediction, and likelihood calculation.
However, the dependency on event types introduces many more parameters
to the POHMM than its HMM counterpart. Therefore, we address the problem
of parameter smoothing, which acts as a kind of regularization and
avoids overfitting \citep{SAMANTA20143614}. In doing so, we derive
explicit marginal distributions, with event type marginalized out,
and demonstrate the equivalence between the marginalized POHMM and
the standard HMM. The marginal distributions conveniently act as a
kind of backoff, or fallback, mechanism in case of missing data, a
technique rooted in linguistics \citep{jurafsky2000speech}.

The rest of this article is organized as follows. Section \ref{sec:Keystroke-dynamics}
briefly describes keystroke dynamics as a behavioral biometric. Section
\ref{sec:pohmm} introduces the POHMM, followed by a simulation study
in Section \ref{sec:simulation} and a case study of the POHMM applied
to keystroke dynamics in Section \ref{sec:case-study}. Section \ref{sec:discussion}
reviews previous modeling efforts for latent processes with partial
observability and contains a discussion. Finally, Section \ref{sec:Conclusion}
concludes the article. The POHMM is implemented in the \texttt{pohmm}
Python package and source code is publicly available\footnote{Available at \href{https://github.com/vmonaco/pohmm}{https://github.com/vmonaco/pohmm}
and through PyPI.}.

\section{Keystroke dynamics \label{sec:Keystroke-dynamics}}

Keystroke dynamics refers to the way a person types. Prominently,
this includes the timings of key press and release events, where each
keystroke is comprised of a press time $t_{n}$ and a duration $d_{n}$.
The time interval between key presses, $\tau_{n}=t_{n}-t_{n-1}$,
is of interest. Compared to random time intervals (RTIs) in which
a user presses only a single key \citep{laskaris2009use}, key press
time intervals occur between different keys and are thought to be
dependent on key distance \citep{salthouse1986perceptual}. A user's
keystroke dynamics is also thought to be relatively unique to the
user, which enable biometric applications, such as user identification
and verification \citep{banerjee2012biometric}.

As a behavioral biometric, keystroke dynamics enables low-cost and
non-intrusive user identification and verification. Keystroke dynamics-based
verification can be deployed remotely, often as a second factor to
username-password verification. Some of the same attributes that make
keystroke dynamics attractive as a behavioral biometric also present
privacy concerns \citep{monaco2016obfuscating}, as there exist numerous
methods of detecting keystrokes without running software on the victim's
computer. Recently, it has been demonstrated that keystrokes can be
detected through a wide range of modalities including motion \citep{wang2015mole},
acoustics \citep{asonov2004keyboard}, network traffic \citep{song2001timing},
and even WiFi signal distortion \citep{ali2015keystroke}.

Due to the keyboard being one of the primary human-computer interfaces,
it is also natural to consider keystroke dynamics as a modality for
\emph{continuous verification} in which a verification decision is
made upon each key pressed throughout a session \citep{bours2015performance}.
Continuous verification holds the promise of greater security, as
users are verified continuously throughout a session beyond the initial
login, which is considered a form of \emph{static verification}. Being
a sequential model, the POHMM is straightforward to use for continuous
verification in addition to identification and static verification.

Keystroke time intervals emanate from a combination of physiology
(e.g., age, gender, and handedness \citep{idrus2014soft}), motor
behavior (e.g., typing skill \citep{salthouse1986perceptual}), and
higher-level cognitive processes \citep{brizan2015utilizing}, highlighting
the difficulty in capturing a user's typing behavior in a succinct
model. Typing behavior generally evolves over time, with highly-practiced
sequences able to be typed much quicker \citep{MONTALVAO201580}.
In biometrics, this is referred to as \emph{template aging}. A user's
keystroke dynamics is also generally dependent on the typing task.
For example, the time intervals observed during password entry are
much different than those observed during email composition.

\section{Partially observable hidden Markov model\label{sec:pohmm}}

The POHMM is intended for applications in which a sequence of \emph{event
types} provides context for an observed sequence of \emph{time intervals}.
This reasoning extends activities other than keystroke dynamics, such
as email, in which a user might be more likely to take an extended
break after sending an email instead of receiving an email, and programming,
in which a user may fix bugs quicker than making feature additions.
The events types form an independent Markov chain and are observed
alongside the sequence of time intervals. This is in contrast to HMM
variants where the neighboring observations themselves provide a context,
such as the adjacent characters in a handwritten segment \citep{bianne2011dynamic}.
Instead, the event types are independent of the dynamics of the model.

With this structure, a distinction can be made between user \emph{behavior}
and \emph{task}: the time intervals comprise the \emph{behavior},
and the sequence of event types, (e.g., the keys pressed) comprise
the \emph{task.} While the time intervals reflect \emph{how} the user
behaves, the sequence of events characterize \emph{what} the user
is doing. This distinction is appropriate for keystroke dynamics,
in which the aim is to capture typing behavior but not the text itself
which may more appropriately modeled by linguistic analysis. Alternatively,
in case the user transcribes a sequence, such as in typing a password,
the task is clearly defined, i.e. the user is instructed to type a
particular sequence of characters. The POHMM aims to capture the temporal
behavior, which depends on the task.

\subsection{Description}

\begin{figure}[t]
\begin{centering}
\includegraphics[width=0.6\columnwidth]{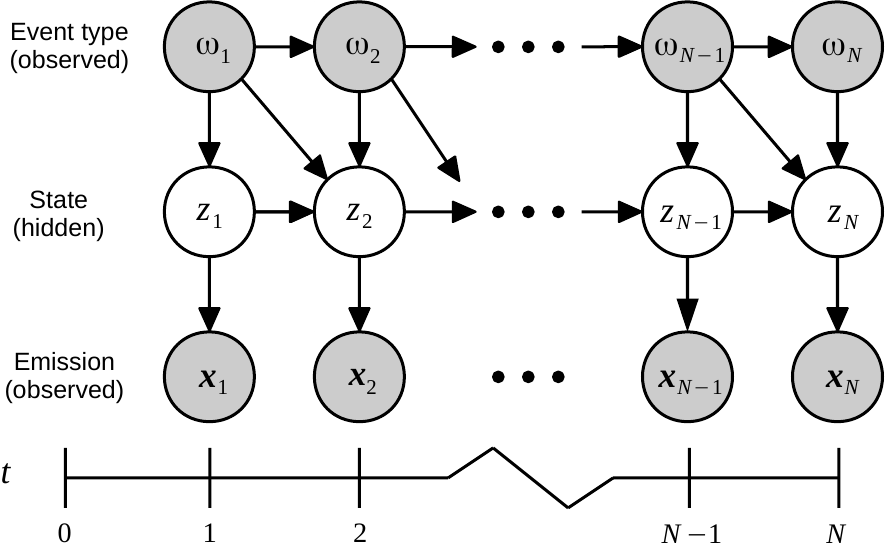} 
\par\end{centering}
\caption{Partially observable hidden Markov model structure. Observed values
(emission and event type) are shown in gray, hidden values (system
state) are shown in white.\label{fig:pohmm-structure}}
\end{figure}

The HMM is a finite-state model in which observed values at time $t$
depend on an underlying latent process \citep{rabiner1989tutorial}.
At the $n^{\mathrm{th}}$ time step $t_{n}$, a feature vector ${\bf x}_{n}$
is emitted and the system can be in any one of $M$ hidden states,
$z_{n}$. Let ${\bf x}_{1}^{N}$ be the sequence of observed emission
vectors and $z_{1}^{N}$ the sequence of hidden states, where $N$
is the total number of observations. The basic HMM is defined by the
recurrence relation,

\begin{equation}
P\left({\bf x}_{1}^{n+1},z_{1}^{n+1}\right)=P\left({\bf x}_{1}^{n},z_{1}^{n}\right)P\left({\bf x}_{n+1}|z_{n+1}\right)P\left(z_{n+1}|z_{n}\right)\;.\label{eq:hmm-axiom}
\end{equation}
The POHMM is an extension of the HMM in which the hidden state and
emission depend on an observed independent Markov chain. Starting
with the HMM axiom in Equation \ref{eq:hmm-axiom}, the POHMM is derived
through following assumptions:
\begin{enumerate}
\item An independent Markov chain of event types is given, denoted by $\Omega_{1}^{N}$.
\item The emission ${\bf x}_{n+1}$ depends on event type $\Omega_{n+1}$
in addition to $z_{n+1}$.
\item The hidden state $z_{n+1}$ depends on $\Omega_{n}$ and $\Omega_{n+1}$
in addition to $z_{n}$.
\end{enumerate}
Applying the above assumptions to the HMM axiom, the conditional emission
probability $P\left({\bf x}_{n+1}|z_{n+1}\right)$ becomes $P\left({\bf x}_{n+1}|z_{n+1},\Omega_{n+1}\right)$;
the conditional hidden state probability $P\left(z_{n+1}|z_{n}\right)$
becomes $P\left(z_{n+1}|z_{n},\Omega_{n},\Omega_{n+1}\right)$; and
the recurrence relation still holds. The complete POHMM axiom is given
by the formula, 
\begin{equation}
P\left({\bf x}_{1}^{n+1},z_{1}^{n+1}\right)=P\left({\bf x}_{1}^{n},z_{1}^{n}\right)P\left({\bf x}_{n+1}|z_{n+1},\Omega_{n+1}\right)P\left(z_{n+1}|z_{n},\Omega_{n},\Omega_{n+1}\right)\label{eq:pohmm-axiom}
\end{equation}
where $\Omega_{n}$ and $\Omega_{n+1}$ are the observed event types
at times $t_{n}$ and $t_{n+1}$. The POHMM structure is shown in
Figure \ref{fig:pohmm-structure}.

The event types come from a finite alphabet of size $m$. Thus, while
the HMM has $M$ hidden states, a POHMM with $m$ event types has
$M$ hidden states per event type, for a total of $m\times M$ unique
hidden states. 

The event type can be viewed as a partial indexing to a much larger
state space. Each observed event type restricts the model to a particular
subset of $M$ hidden states with differing probabilities of being
in each hidden state, hence the partial observability. The POHMM starting
and emission probabilities can be viewed as an HMM for each event
type, and the POHMM transition probabilities as an HMM for each \emph{pair}
of event types.

\begin{figure}[t]
\begin{centering}
\includegraphics[width=0.6\columnwidth]{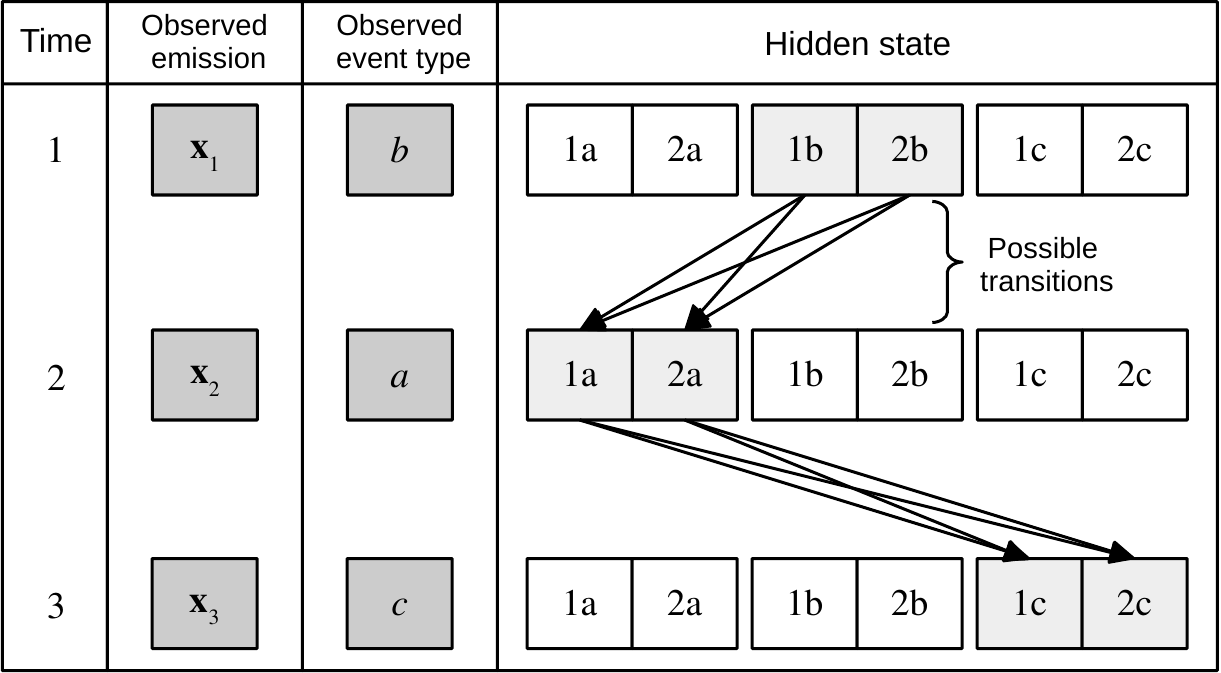} 
\par\end{centering}
\caption{POHMM event types index a much larger state space. In this example,
there are two hidden states and three event types. Given observed
event type $b$ at time $1$, the system must be in one of the hidden
states $\left\{ 1b,2b\right\} $. The $a$ observed at the next time
step limits the possible transitions from $\left\{ 1b,2b\right\} $
to $\left\{ 1a,2a\right\} $.\label{fig:pohmm-hidden-state-example}}
\end{figure}

To illustrate this concept, consider a POHMM with two hidden states
and three event types, where $\Omega_{1}^{3}=\left[b,a,c\right]$.
At each time step, the observed event type limits the system to hidden
states that have been conditioned on that event type, as demonstrated
in Figure \ref{fig:pohmm-hidden-state-example}. Beginning at time
$1$, given observed event type $\Omega_{1}=b$, the system must be
in one of the hidden states $\left\{ 1b,2b\right\} $. Event type
$\Omega_{2}=a$ observed at time $2$ then restricts the possible
transitions from $\left\{ 1b,2b\right\} $ to $\left\{ 1a,2a\right\} $.
Generally, given any event type, the POHMM must be in one of $M$
hidden states conditioned on that event type. Section \ref{subsec:Marginal-distributions}
deals with situations where the event type is missing or has not been
previously observed in which case the marginal distributions (with
the event type marginalized out) are used.

The POHMM parameters are derived from the HMM. Model parameters include
$\pi\left[j|\omega\right]$, the probability of starting in state
$j$ given event type $\omega$, and $a\left[i,j|\psi,\omega\right]$,
the probability of transitioning from state $i$ to state $j$, given
event types $\psi$ and $\omega$ before and after the transition,
respectively\footnote{When a transition is involved, $i$ and $\psi$ always refer to the
hidden state and event type, respectively, before the transition;
$j$ and $\omega$ refer to those after the transition.}. Let $f\left(\cdot;{\bf b}\left[j|\omega\right]\right)$ be the emission
distribution that depends on hidden state $j$ and event type $\omega$,
where ${\bf b}\left[j|\omega\right]$ parametrizes density function
$f\left(\cdot\right)$. The complete set of parameters is denoted
by $\theta=\left\{ \pi,a,{\bf b}\right\} $, where $a$ is the $m^{2}M^{2}$
transition matrix. While the total number of parameters in the HMM
is $M+M^{2}+MK$, where $K$ is the number of free parameters in the
emission distribution, the POHMM contains $mM+m^{2}M^{2}+mMK$ parameters.
After accounting for normalization constraints, the degrees of freedom
(\emph{dof}) is $m\left(M-1\right)+m^{2}M\left(M-1\right)+mMK$.

Marginal distributions, in which the event type is marginalized out,
are also defined. Let $\pi\left[j\right]$ and $f\left(\cdot;{\bf b}\left[j\right]\right)$
be the marginalized starting and emission probabilities, respectively.
Similarly, the parameters $a\left[i,j|\omega\right]$, $a\left[i,j|\psi\right]$,
and $a\left[i,j\right]$ are defined as the transition probabilities
after marginalizing out the first, second, and both event types, respectively.
\emph{The POHMM marginal distributions are exactly equal to the corresponding
HMM that ignores the event types}. This ensures that the POHMM is
no worse than the HMM in case the event types provide little or no
information as to the process being modeled. Computation of POHMM
marginal distributions is covered in Section \ref{subsec:Marginal-distributions}
and simulation results demonstrating this equivalence are in Section
\ref{sec:simulation}.

It may seem that POHMM parameter estimation becomes intractable, as
the number of possible transitions between hidden states increases
by a factor of $m^{2}$ over the HMM and all other parameters by a
factor of $m$. In fact, \emph{all of the algorithms used for the
POHMM are natural extensions of those used for the HMM}: the POHMM
parameters and variables are adapted from the HMM by introducing the
dependence on event types, and parameter estimation and likelihood
calculation follow the same basic derivations as those for the HMM.
POHMM parameter estimation remains linearly bounded in the number
of observations, similar to the HMM, performed through a modification
of the Baum-Welch (BW) algorithm. The convergence property of the
modified BW algorithm is demonstrated analytically in Section \ref{sec:Parameter-estimation}
and empirically in Section \ref{sec:simulation}. The rest of this
section addresses the three main problems of the POHMM, taken analogously
as the three main problems of the HMM:
\begin{enumerate}
\item Determine $P({\bf x}_{1}^{N}|\Omega_{1}^{N},\theta)$, the likelihood
of an emission sequence given the model parameters and the observed
event types. 
\item Determine $z_{1}^{N}$, the maximum likelihood sequence of hidden
states, given the emissions ${\bf x}_{1}^{N}$ and event types $\Omega_{1}^{N}$.
\item Determine $\arg\max_{\theta\in\Theta}P({\bf x}_{1}^{N}|\Omega_{1}^{N},\theta)$,
the maximum likelihood parameters $\theta$ for observed emission
sequence ${\bf x}_{1}^{N}$, given the event type sequence.
\end{enumerate}
The first and third problems are necessary for identifying and verifying
users in biometric applications, while the second problem is useful
for understanding user behavior. The rest of this section reviews
the solutions to each of these problems and other aspects of parameter
estimation, including parameter initialization and smoothing.

\subsection{Model likelihood\label{sec:Model-likelihood}}

Since we assume $\Omega_{1}^{N}$ is given, it does not have a prior
distribution. Therefore, we consider only the likelihood of an emission
sequence given the model parameters $\theta$ and the observed event
type sequence $\Omega_{1}^{N}$, denoted by $P({\bf x}_{1}^{N}|\Omega_{1}^{N})$\footnote{For brevity, the dependence on $\theta$ is implied, writing $P({\bf x}_{1}^{N}|\Omega_{1}^{N},\theta)$
as $P({\bf x}_{1}^{N}|\Omega_{1}^{N})$.}, leaving the joint model likelihood $P({\bf x}_{1}^{N},\Omega_{1}^{N})$
as an item for future work.

In the HMM, $P({\bf x}_{1}^{N})$ can be computed efficiently by the
forward procedure which defines a recurrence beginning at the start
of the sequence. This procedure differs slightly for the POHMM due
to the dependence on event types. Notably, the starting, transition,
and emission parameters are all conditioned on the given event type.

Let $\alpha_{n}\left[z_{n},\Omega_{n}\right]\equiv P\left({\bf x}_{1}^{n},z_{n}|\Omega_{n}\right)$,
i.e., the joint probability of emission subsequence ${\bf x}_{1}^{n}$
and hidden state $z_{n}$, given event type $\Omega_{n}$. Then, by
the POHMM axiom (Equation \ref{eq:pohmm-axiom}), $\alpha_{n}\left[z_{n},\Omega_{n}\right]$
can be computed recursively by the formula,

\begin{eqnarray}
\alpha_{n+1}\left[z_{n+1},\Omega_{n+1}\right] & = & P\left({\bf x}_{n+1}|z_{n+1},\Omega_{n+1}\right)\label{eq:pohmm-axiom-forward}\\
 &  & \times\sum_{z_{n}}P\left(z_{n+1}|z_{n},\Omega_{n},\Omega_{n+1}\right)\alpha_{n}\left[z_{n},\Omega_{n}\right]\nonumber \\
\alpha_{1}\left[z_{1},\Omega_{1}\right] & = & P\left({\bf x}_{1}|z_{1},\Omega_{1}\right)P\left(z_{1}|\Omega_{1}\right)\label{eq:pohmm-axiom-forward-base}
\end{eqnarray}
where Equation \ref{eq:pohmm-axiom-forward-base} provides the initial
condition. The modified forward algorithm is obtained by substituting
the model parameters into Equations \ref{eq:pohmm-axiom-forward}
and \ref{eq:pohmm-axiom-forward-base}, where
\begin{eqnarray}
\pi\left[j|\omega\right] & \equiv & P\left(z_{1}=j|\Omega_{1}=\omega\right)\\
f({\bf x}_{n};{\bf b}\left[j,\omega\right]) & \equiv & P\left({\bf x}_{n}|z_{n}=j,\Omega_{n}=\omega\right)\\
a\left[i,j|\psi,\omega\right] & \equiv & P\left(z_{n+1}=j|z_{n}=i,\Omega_{n}=\psi,\Omega_{n+1}=\omega\right)\label{eq:pohmm-parameter-equivs}
\end{eqnarray}
and $\alpha_{n}\left[j,\omega\right]$ is the sequence obtained after
substituting the model parameters. The model likelihood is easily
computed upon termination, since {\small{}$P({\bf x}_{1}^{N}|\Omega_{1}^{N})=\sum_{j=1}^{M}\alpha_{N}\left[j,\omega\right]$
where $\omega=\Omega_{N}$.}{\small \par}

A modified backward procedure is similarly defined through a backwards
recurrence. Let $\beta_{n}\left[z_{n},\Omega_{n}\right]\equiv P\left({\bf x}_{n+1}^{N}|z_{n},\Omega_{n}\right)$.
Then under the POHMM axiom,

\begin{eqnarray}
\beta_{n}\left[z_{n},\Omega_{n}\right] & = & \sum_{z_{n+1}}P\left({\bf x}_{n+1}|z_{n+1},\Omega_{n+1}\right)\label{eq:pohmm-axiom-backward}\\
 &  & \quad\times P\left(z_{n+1}|z_{n},\Omega_{n},\Omega_{n+1}\right)\beta_{n+1}\left[z_{n+1},\Omega_{n+1}\right]\nonumber \\
\beta_{N}\left[z_{N},\Omega_{N}\right] & = & 1\;.
\end{eqnarray}
where $\beta_{n}\left[j,\omega\right]$ is the sequence obtained after
making the same substitutions.

Note that at each $n$, $\alpha_{n}\left[j,\omega\right]$ and $\beta_{n}\left[j,\omega\right]$
need only be computed for the observed $\omega=\Omega_{n}$, i.e.,
we don't care about event types $\omega\ne\Omega_{n}$. Therefore,
only the hidden states (and not the event types) are enumerated in
Equations \ref{eq:pohmm-axiom-forward} and \ref{eq:pohmm-axiom-backward}
at each time step. Like the HMM, the modified forward and backward
algorithms have time complexity $O(M^{2}N)$ and can be stored in
a $N\times M$ matrix.

\subsection{Hidden state prediction\label{sec:Hidden-states}}

The maximum likelihood sequence of hidden states is efficiently computed
using the event type-dependent forward and backward variables defined
above. First, let the POHMM forward-backward variable $\gamma_{n}\left[z_{n},\Omega_{n}\right]\equiv P\left(z_{n}|\Omega_{n},{\bf x}_{1}^{N}\right)$,
i.e., the posterior probability of hidden state $z_{n}$, given event
type $\Omega_{n}$ and the emission sequence ${\bf x}_{1}^{N}$. Let
$\gamma_{n}\left[j,\omega\right]$ be the estimate obtained using
the model parameters, making the same substitutions as above. Then
$\gamma_{n}\left[j,\omega\right]$ is straightforward to compute using
the forward and backward variables, given by
\begin{eqnarray}
\gamma_{n}\left[j,\omega\right] & = & \frac{\alpha_{n}\left[j|\omega\right]\beta_{n}\left[j|\omega\right]}{P({\bf x}_{1}^{N}|\Omega_{1}^{N})}\label{eq:pohmm-forward-backward-pohmm}\\
 & = & \frac{\alpha_{n}\left[j|\omega\right]\beta_{n}\left[j|\omega\right]}{\sum_{i=1}^{M}\alpha_{n}\left[i|\omega\right]\beta_{n}\left[i|\omega\right]}\nonumber 
\end{eqnarray}
where $\omega=\Omega_{n}$. The sequence of maximum likelihood hidden
states is taken as,
\begin{equation}
z_{n}=\mbox{arg max}_{1\le j\le M}\gamma_{n}\left[j,\omega\right]\;.\label{eq:pohmm-most-likely-hidden-states}
\end{equation}
Similar to $\alpha_{n}\left[j|\omega\right]$ and $\beta_{n}\left[j|\omega\right]$,
$\gamma_{n}\left[j,\omega\right]$ can be stored in a $N\times M$
matrix and takes $O\left(M^{2}N\right)$ time to compute. This is
due to the fact that the event types are not enumerated at each step;
the dependency on the event type propagates all the way to the re-estimated
parameters, defined below.

\subsection{Parameter estimation\label{sec:Parameter-estimation}}

\begin{algorithm}
\begin{enumerate}
\item \textbf{\small{}Initialization}{\small{}}\\
{\small{} Choose initial parameters $\theta_{o}$ and let $\theta\leftarrow\theta_{o}$.}{\small \par}
\item \textbf{\small{}Expectation}{\small{}}\\
{\small{} Use $\theta$, ${\bf x}_{1}^{N}$, $\Omega_{1}^{N}$ to
compute $\alpha_{n}\left[j|\omega\right]$, $\beta_{n}\left[j|\omega\right],$
$\gamma_{n}\left[j,\omega\right]$, $\xi_{n}\left[i,j|\psi,\omega\right]$.}{\small \par}
\item \textbf{\small{}Maximization}{\small{}}\\
{\small{} Update $\theta$ using the re-estimation formulae (Eqs.
\ref{eq:update-startproba-pohmm}, \ref{eq:update-transmat-pohmm},
\ref{eq:update-emission-pohmm}) to get $\dot{\theta}=\left\{ \dot{\pi},\dot{a},\dot{{\bf b}}\right\} $. }{\small \par}
\item \textbf{\small{}Regularization}{\small{}}\\
{\small{} Calculate marginal distributions and apply parameter smoothing
formulae.}{\small \par}
\item \textbf{\small{}Termination}{\small{}}\\
{\small{} If $\ln P\left({\bf x}_{1}^{N}|\Omega_{1}^{N},\dot{\theta}\right)-\ln P\left({\bf x}_{1}^{N}|\Omega_{1}^{N},\theta\right)<\epsilon$,
stop; else let $\theta\leftarrow\dot{\theta}$ and go to step 2.}{\small \par}
\end{enumerate}
\caption{Modified Baum-Welch for POHMM parameter estimation.\label{alg:pohmm-baum-welch-algorithm-with-smoothing}}
\end{algorithm}

Parameter estimation is performed iteratively, updating the starting,
transition, and emission parameters using the current model parameters
and observed sequences. In each iteration of the modified Baum-Welch
algorithm, summarized in Algorithm \ref{alg:pohmm-baum-welch-algorithm-with-smoothing},
the model parameters are re-estimated using the POHMM forward, backward,
and forward-backward variables. Parameters are set to initial values
before the first iteration, and convergence is reached upon a loglikelihood
increase of less than $\epsilon$.

\subsubsection{Starting parameters}

Using the modified forward-backward variable given by Equation \ref{eq:pohmm-forward-backward-pohmm},
the re-estimated POHMM starting probabilities are obtained directly
by
\begin{equation}
\dot{\pi}\left[j|\omega\right]=\gamma_{1}\left[j|\omega\right]\label{eq:update-startproba-pohmm}
\end{equation}
where $\omega=\Omega_{1}$ and re-estimated parameters are denoted
by a dot. Generally, it may not be possible to estimate $\dot{\pi}\left[j|\omega\right]$
for many $\omega$ due to there only being one $\Omega_{1}$ (or several
$\Omega_{1}$ for multiple observation sequences). Parameter smoothing,
introduced in Section \ref{subsec:smoothing-parameters}, addresses
this issue of missing and infrequent observations.

\subsubsection{Transition parameters}

In contrast to the HMM, which has $M^{2}$ transition probabilities,
there are $m^{2}M^{2}$ unique transition probabilities in the POHMM.
Let $\xi_{n}\left[z_{n},z_{n+1}|\Omega_{n},\Omega_{n+1}\right]\equiv P\left(z_{n+1}|z_{n},\Omega_{n},\Omega_{n+1},{\bf x}_{1}^{N}\right)$,
i.e., the probability of transitioning from state $z_{n}$ to $z_{n+1}$,
given event types $\Omega_{n}$ and $\Omega_{n+1}$ as well as the
emission sequence. Substituting the forward and backward variable
estimates based on model parameters, this becomes $\xi_{n}\left[i,j|\psi,\omega\right]$,
given by 
\begin{equation}
\xi_{n}\left[i,j|\psi,\omega\right]=\frac{\alpha_{n}\left[i,\omega\right]a\left[i,j|\psi,\omega\right]f\left({\bf x}_{n+1};{\bf b}\left[j|\omega\right]\right)\beta_{n}\left[j|\omega\right]}{P({\bf x}_{1}^{N}|\Omega_{1}^{N})}\;.\label{eq:expectation-xi-pohmm}
\end{equation}
for $1\le n\le N-1$, $\psi=\Omega_{n}$ and $\omega=\Omega_{n+1}$.
The updated transition parameters are then calculated by 
\begin{equation}
\dot{a}\left[i,j|\psi,\omega\right]=\frac{\sum_{n=1}^{N-1}\xi_{n}\left[i,j|\psi,\omega\right]\delta\left(\psi,\Omega_{n}\right)\delta\left(\omega,\Omega_{n+1}\right)}{\sum_{n=1}^{N-1}\gamma_{n}\left[i|\psi\right]\delta\left(\psi,\Omega_{n}\right)\delta\left(\omega,\Omega_{n+1}\right)}\label{eq:update-transmat-pohmm}
\end{equation}
where $\delta\left(\omega,\Omega_{n}\right)=1$ if $\omega=\Omega_{n}$
and 0 otherwise. Note that $\dot{a}\left[i,j|\psi,\omega\right]$
depends only on the transitions between event types $\psi$ and $\omega$
in $\Omega_{1}^{N}$, i.e., where $\Omega_{n}=\psi$ and $\Omega_{n+1}=\omega$,
as the summand in the numerator equals $0$ otherwise. As a result,
the updated transition probabilities can be computed in $O(M^{2}N)$
time, the same as the HMM, despite there being $m^{2}M^{2}$ unique
transitions.

\subsubsection{Emission parameters}

For each hidden state and event type, the emission distribution parameters
are re-estimated through the optimization problem,
\begin{equation}
\dot{{\bf b}}\left[j|\omega\right]=\arg\max_{{\bf b}\in\mathcal{B}}\sum_{n=1}^{N}\gamma_{n}\left[j|\omega\right]\ln f\left({\bf x}_{n};{\bf b}\right)\delta\left(\omega,\Omega_{n}\right)\;.\label{eq:update-emission-pohmm}
\end{equation}
Closed-form expressions exist for a variety of emission distributions.
In this work, we use the log-normal density for time intervals. The
log-normal has previously been demonstrated as a strong candidate
for modeling keystroke time intervals, which resemble a heavy-tailed
distribution \citep{MONTALVAOFILHO20061440}. The log-normal density
is given by

\begin{equation}
f(x;\eta,\rho)=\frac{1}{x\rho\sqrt{2\pi}}\exp\left[\frac{-\left(\ln x-\eta\right)^{2}}{2\rho^{2}}\right]\label{eq:log-normal}
\end{equation}
where $\eta$ and $\rho$ are the log-mean and log-standard deviation,
respectively. The emission parameter re-estimates are given by 
\begin{equation}
{\bf \dot{\eta}}\left[j|\omega\right]=\frac{\sum_{n=1}^{N}\gamma_{n}\left[j|\omega\right]\ln\tau_{n}\delta\left(\omega,\Omega_{n}\right)}{\sum_{n=1}^{N}\gamma_{n}\left[j|\omega\right]\delta\left(\psi,\Omega_{n}\right)}\label{eq:update-mean-pohmm}
\end{equation}
and 
\begin{equation}
{\bf \dot{\rho}}^{2}\left[j|\omega\right]=\frac{\sum_{n=1}^{N}\gamma_{n}\left[j|\omega\right](\ln\tau_{n}-\dot{\eta}_{j|\omega})^{2}\delta\left(\omega,\Omega_{n}\right)}{\sum_{n=1}^{N}\gamma_{n}\left[j|\omega\right]\delta\left(\psi,\Omega_{n}\right)}\label{eq:update-variance-pohmm}
\end{equation}
for hidden state $j$, given event type $\omega$. Note that the estimates
for ${\bf \dot{\eta}}\left[j|\omega\right]$ and ${\bf \dot{\rho}}\left[j|\omega\right]$
depend only on the elements of $\gamma_{n}\left[j|\omega\right]$
where $\Omega_{n}=\omega$.

\subsubsection{Convergence properties}

The modified Baum-Welch algorithm for POHMM parameter estimation (Algorithm
\ref{alg:pohmm-baum-welch-algorithm-with-smoothing}) relies on the
principles of expectation maximization (EM) and is guaranteed to converge
to a local maximum. The re-estimation formula (Equations \ref{eq:update-startproba-pohmm},
\ref{eq:update-transmat-pohmm}, and \ref{eq:update-emission-pohmm})
are derived from inserting the model parameters from two successive
iterations, $\theta$ and $\dot{\theta}$, into Baum's auxiliary function,
$Q\left(\theta,\dot{\theta}\right)$, and maximizing $Q\left(\theta,\dot{\theta}\right)$
with respect to the updated parameters. Convergence properties are
evaluated empirically in Section \ref{sec:simulation}, and \ref{sec:Proof-of-convergence}
contains a proof of convergence, which follows that of the HMM.

\subsection{Parameter initialization\label{subsec:Parameter-initialization}}

Parameter estimation begins with parameter initialization, which plays
an important role in the BW algorithm and may ultimately determine
the quality of the estimated model since EM guarantees only locally
maximum likelihood estimates. This work uses an observation-based
parameter initialization procedure that ensures reproducible parameter
estimates, as opposed to random initialization. The starting and transition
probabilities are simply initialized as
\begin{eqnarray}
\pi\left[j|\omega\right] & = & \frac{1}{M}\label{eq:initialize-start-prob}\\
a\left[i,j|\psi,\omega\right] & = & \frac{1}{M}\label{eq:initialize-transition-prob}
\end{eqnarray}
for all $i$, $j$, $\psi$, and $\omega$. This reflects maximum
entropy, i.e., uniform distribution, in the absence of any starting
or transition priors.

Next, the emission distribution parameters are initialized. The strategy
proposed here is to initialize parameters in such a way that there
is a correspondence between hidden states from two different models.
That is, for any two models with $M=2$, hidden state $j=1$ corresponds
to the active state and $j=2$ corresponds to the passive state. Using
a log-normal emission distribution, this is accomplished by spreading
the log-mean initial parameters. Let
\begin{equation}
{\bf \eta}\left[\omega\right]=\frac{\sum_{n=1}^{N}\ln{\bf x}_{n}\delta\left(\omega,\Omega_{n}\right)}{\sum_{n=1}^{N}\delta\left(\omega,\Omega_{n}\right)}\label{eq:observation-emission-mean}
\end{equation}
and
\begin{equation}
\rho^{2}\left[\omega\right]=\frac{\sum_{n=1}^{N}(\ln{\bf x}_{n}-\eta\left[\omega\right])^{2}\delta\left(\omega,\Omega_{n}\right)}{\sum_{n=1}^{N}\delta\left(\omega,\Omega_{n}\right)}\label{eq:observation-emission-std}
\end{equation}
be the observed log-mean and log-variance for event type $\omega$.
The model parameters are then initialized as
\begin{equation}
{\bf \eta}\left[j|\omega\right]=\eta\left[\omega\right]+\left(\frac{2h\left(j-1\right)}{M-1}-h\right)\rho\left[\omega\right]\label{eq:initialize-emission-mean}
\end{equation}
and
\begin{equation}
\rho^{2}\left[j|\omega\right]=\rho^{2}\left[\omega\right]\label{eq:initialize-emission-std}
\end{equation}
for $1\le j\le M$, where $h$ is a bandwidth parameter. Using $h=2,$
initial states are spread over the interval $\left[\eta\left[\omega\right]-2\rho\left[\omega\right],\eta\left[\omega\right]+2\rho\left[\omega\right]\right]$,
i.e., 2 log-standard deviations around the log-mean. This ensures
that $j=1$ corresponds to the state with the smaller log-mean, i.e.,
the active state.

\subsection{Marginal distributions\label{subsec:Marginal-distributions}}

When computing the likelihood of a novel sequence, it is possible
that some event types were not encountered during parameter estimation.
This situation arises when event types correspond to key names of
freely-typed text and novel key sequences are observed during testing.
A fallback mechanism (sometimes referred to as a ``backoff'' model)
is typically employed to handle missing or sparse data, such as that
used linguistics \citep{jurafsky2000speech}. In order for the POHMM
to handle missing or novel event types during likelihood calculation,
the marginal distributions are used. This creates a two-level fallback
hierarchy in which missing or novel event types fall back to the distribution
in which the event type is marginalized out.

Note also that while we assume $\Omega_{1}^{N}$ is given (i.e., has
no prior), the individual $\Omega_{n}$ \emph{do} have a prior defined
by their occurrence in $\Omega_{1}^{N}$. It is this feature that
enables the event type to be marginalized out to obtain the equivalent
HMM. Let the probability of event type $\omega$ at time $t_{1}$
be $\pi\left[\omega\right]$, and the probability of transitioning
from event type $\psi$ to $\omega$ be denoted by $a\left[\psi,\omega\right]$.
Both can be computed directly from the event type sequence $\Omega_{1}^{N}$,
which is assumed to be a first-order Markov chain. The marginal $\pi\left[j\right]$
is the probability of starting in hidden state $j$ in which the event
type has been marginalized out,
\begin{equation}
\pi\left[j\right]=\sum_{\omega\in\Omega}\pi\left[j|\omega\right]\pi\left[\omega\right]\label{eq:pohmm-startprob-marginal}
\end{equation}
where $\Omega$ is the set of unique event types in $\Omega_{1}^{N}$.

Marginal transition probabilities are also be defined. Let $a\left[i,j|\psi\right]$
be the probability of transitioning from hidden state $i$ to hidden
state $j$, given event type $\psi$ while in hidden state $i$. The
second event type for hidden state $j$ has been marginalized out.
This probability is given by
\begin{equation}
a\left[i,j|\psi\right]=\sum_{\omega\in\Omega}a\left[i,j|\psi,\omega\right]a\left[\psi,\omega\right]\;.\label{eq:pohmm-marginal-transmat-aij-omega-dot}
\end{equation}
The marginal probability $a\left[i,j|\omega\right]$ is defined similarly
by
\begin{equation}
a\left[i,j|\omega\right]=\frac{\sum_{\psi\in\Omega}a\left[i,j|\psi,\omega\right]a\left[\psi,\omega\right]}{\sum_{\psi\in\Omega}a\left[\psi,\omega\right]}\;.\label{eq:pohmm-marginal-transmat-aji-dot-omega}
\end{equation}
Finally, the marginal $a\left[i,j\right]$ is the probability of transitioning
from $i$ to $j$,
\begin{equation}
a\left[i,j\right]=\frac{1}{m}\sum_{\psi\in\Omega}\sum_{\omega\in\Omega}a\left[i,j|\psi,\omega\right]a\left[\psi,\omega\right]\;.\label{eq:pohmm-marginal-transmat}
\end{equation}
No denominator is needed in Equation \ref{eq:pohmm-marginal-transmat-aij-omega-dot}
since the normalization constraints of both transition matrices carry
over to the left-hand side. Equation \ref{eq:pohmm-marginal-transmat}
is normalized by $\frac{1}{m}$ since $\sum_{\psi\in\Omega}\sum_{\omega\in\Omega}a\left[\psi,\omega\right]=m$.

The marginal emission distribution is a convex combination of the
emission distributions conditioned on each of the event types. For
normal and log-normal emissions, the marginal emission is simply a
mixture of normals or log-normals, respectively. Let $\eta\left[j\right]$
and $\rho^{2}\left[j\right]$ be the log-mean and log-variance of
the marginal distribution for hidden state $j$. The marginal log-mean
is a weighted sum of the conditional distributions, given by
\begin{equation}
\eta\left[j\right]=\sum_{\omega\in\Omega}\Pi\left[\omega\right]\mu\left[j|\omega\right]\label{eq:mixture-mean}
\end{equation}
where $\Pi\left[\omega\right]$ is the stationary probability of event
type $\omega$. This can be calculated directly from the event type
sequence $\Omega_{1}^{N}$, 
\begin{equation}
\Pi\left[\omega\right]=\frac{1}{N}\sum_{n=1}^{N}\delta\left(\omega,\Omega_{n}\right)\;.\label{eq:statestate-proba-given-pstate}
\end{equation}
Similarly, the marginal log-variance is a mixture of log-normals given
by
\begin{equation}
\rho^{2}\left[j\right]=\sum_{\omega\in\Omega}\Pi\left[\omega\right]\left[\left(\eta\left[j|\omega\right]-\eta\left[j\right]\right)^{2}+\rho^{2}\left[j|\omega\right]\right]\;.\label{eq:mixture-covariance}
\end{equation}
Marginalized distribution parameters for normal emission is exactly
the same.

\subsection{Parameter smoothing\label{subsec:smoothing-parameters}}

HMMs with many hidden states (and parametric models in general) are
plagued by overfitting and poor generalization, especially when the
sample size is small. This has to due with there being a high \emph{dof}
in the model compared to the number of observations. Previous attempts
at HMM parameter smoothing have pushed the emission and transition
parameters towards a higher entropy distribution \citep{SAMANTA20143614}
or borrowed the shape of the emission PDF from states that appear
in a similar context \citep{lee1989speaker}. Instead, our parameter
smoothing approach uses the marginal distributions, which can be estimated
with higher confidence due to there being more observations, to eliminate
the sparseness in the event type-dependent parameters. Note that parameter
smoothing goes hand-in-hand with context-dependent models, at least
in part due to the curse of dimensionality which is introduced by
the context dependence \citep{lee1989speaker}.

The purpose of parameter smoothing is twofold. First, it acts as a
kind of regularization to avoid overfitting, a problem often encountered
when there is a large number of parameters and small number of observations.
Second, parameter smoothing provides superior estimates in case of
missing or infrequent data. For motivation, consider a keystroke sequence
of length $N$. Including English letters and the Space key, there
are at most $27$ unique keys and $729$ unique digrams (subsequences
of length 2). Most of these will rarely, or never, be observed in
a sequence of English text. Parameter smoothing addresses this issue
by re-estimating the parameters that depend on low-frequency observations
using a mixture of the marginal distribution. The effect is to bias
parameters that depend on event types with low frequency toward the
marginals, for which there exist more observations and higher confidence,
while parameters that depend on event types with high frequency will
remain unchanged.

Smoothing weights for the starting and emission parameters are defined
as
\begin{equation}
w_{\omega}=1-\frac{1}{1+f\left(\omega\right)}\label{eq:inverse-frequency-weights}
\end{equation}
where $f(\omega)=\sum_{t=1}^{N}\delta\left(\omega,\Omega_{n}\right)$
is the frequency of event type $\omega$ in the sequence $\Omega_{1}^{N}$.
The POHMM starting probabilities are then smoothed by
\begin{equation}
\tilde{\pi}\left[j|\omega\right]=w_{\omega}\pi\left[j|\omega\right]+\left(1-w_{\omega}\right)\pi\left[j\right]\label{eq:starting-smoothing}
\end{equation}
where smoothed parameter estimates denoted by a tilde, and emission
parameters are smoothed by 
\begin{equation}
\tilde{{\bf b}}\left[j|\omega\right]=w_{\omega}{\bf b}\left[j|\omega\right]+\left(1-w_{\omega}\right){\bf b}\left[j\right]\;.\label{eq:emission-smoothing}
\end{equation}
As $N$ increases, event type frequencies increase and the effect
of parameter smoothing is diminished, while parameters conditioned
on infrequent or missing event types are biased toward the marginal.
This ensures that the conditional parameters remain asymptotically
unbiased as $N\rightarrow\infty$. 

The smoothing weights for transition probabilities follow similar
formulae. Let $f\left(\psi,\omega\right)=\sum_{t=1}^{N-1}\delta\left(\psi,\Omega_{n}\right)\delta\left(\omega,\Omega_{n+1}\right)$,
i.e., the frequency of event type $\psi$ followed by $\omega$ in
the sequence $\Omega_{1}^{N}$. Weights for the conditional and marginal
transition probabilities are defined as
\begin{eqnarray}
w_{\psi} & = & \frac{1}{f\left(\psi,\omega\right)+f\left(\omega\right)}\nonumber \\
w_{\omega} & = & \frac{1}{f\left(\psi,\omega\right)+f\left(\psi\right)}\nonumber \\
w_{\psi,\omega} & = & 1-\left(w_{\psi}+w_{\omega}\right)\nonumber \\
w & = & 0\label{eq:transition-smoothing}
\end{eqnarray}
where $w_{\psi,\omega}+w_{\psi}+w_{\omega}+w=1$. The smoothed transition
matrix is given by
\begin{equation}
\tilde{a}\left[i,j|\psi,\omega\right]=w_{\psi,\omega}a\left[i,j|\psi,\omega\right]+w_{\psi}a\left[i,j|\psi\right]+w_{\omega}a\left[i,j|\omega\right]+wa\left[i,j\right]\;.\label{eq:smoothed-transition-matrix}
\end{equation}
In this strategy, the weight for the marginal $a\left[i,j\right]$
is 0, although in other weighting schemes, $w$ could be non-zero.

\section{Simulation study\label{sec:simulation}}

It is important for statistical models and their implementations to
be consistent. This requires that parameter estimation be both convergent
and asymptotically unbiased. The POHMM \emph{algorithms} include the
parameter estimation procedure and equations, and the \emph{implementation}
consists of the POHMM algorithms expressed in a programming language.
While consistency of the POHMM algorithms is theoretically guaranteed
(proof in \ref{sec:Proof-of-convergence}), consistency of the POHMM
implementation under several different scenarios is validated in this
section using computational methods. 

First, a model is initialized with parameters $\theta_{o}$. From
this model, $S$ samples are generated, each containing $N$ time
intervals. For each sample, the best-estimate parameters $\hat{\theta}$
are computed using the modified BW algorithm (Algorithm \ref{alg:pohmm-baum-welch-algorithm-with-smoothing}).
Let $\hat{\theta}_{N}$ be the parameters determined by the modified
BW algorithm for an observed sequence of length $N$ generated from
a POHMM with true parameters $\theta_{o}$. Consistency requires that
\begin{equation}
\lim_{N\to\infty}\frac{|\hat{\theta}_{N}-\theta_{o}|}{\max_{\hat{\theta}}|\hat{\theta}_{N}-\theta_{o}|}={\bf 0}\label{eq:consistency}
\end{equation}
insensitive to the choice of $\theta_{o}$. As $N$ increases, parameter
estimation should be able to recover the true model parameters from
the observed data. Four different scenarios are considered:
\begin{enumerate}
\item Train a POHMM (without smoothing) on POHMM-generated data.
\item Train a POHMM (with smoothing) on POHMM-generated data.
\item Train a POHMM (without smoothing) using emissions generated from an
HMM and random event types.
\item Train an HMM using emissions from a POHMM (ignore event types).
\end{enumerate}
Convergence is theoretically guaranteed for scenarios 1 and 2. The
first scenario tests the POHMM implementation without parameter smoothing
and should yield unbiased estimates. Scenario 2 evaluates the POHMM
implementation with parameter smoothing, whose effect diminishes as
$N$ increases. Consequently, the smoothed POHMM estimates approach
that of the unsmoothed POHMM, and results should also indicate consistency.

Scenario 3 is a POHMM trained on an HMM, and scenario 4 is an HMM
trained on a POHMM. In scenario 3, the underlying process is an HMM
with the same number of hidden states as the POHMM, and the observed
event types are completely decorrelated from the HMM. As a result,
the event types do \emph{not} partially reveal the hidden state. In
this case, the POHMM marginal distributions, in which the event type
is marginalized out, should converge to the HMM. Finally, scenario
4 simply demonstrates the inability of the HMM to capture the dependence
on event types, and results should indicate biased estimates.

For scenarios 1, 2 and 4, a POHMM with 3 event types and 2 hidden
states is initialized to generate the training data. The emission
distribution is a univariate Gaussian with parameters chosen to be
comparable to human key-press time intervals, and transition probabilities
are uniformly distributed. The emission and event type sequences are
sampled from the POHMM and used to fit the model. In scenario 3, an
HMM generates the emission sequence ${\bf x}_{1}^{N}$, and the event
type sequence $\Omega_{1}^{N}$ is chosen randomly from the set of
3 event types, reflecting no dependence on event types. In this case,
only the POHMM marginal distribution parameter residuals are evaluated,
as these should approximate the underlying HMM. For each value of
$N$ in each scenario, 400 length-$N$ samples are generated and used
to train the corresponding model.

\begin{figure}
\begin{centering}
\subfloat[Studentized emission residuals.\label{fig:Studentized-residuals.}]{\begin{centering}
\includegraphics[width=0.7\columnwidth]{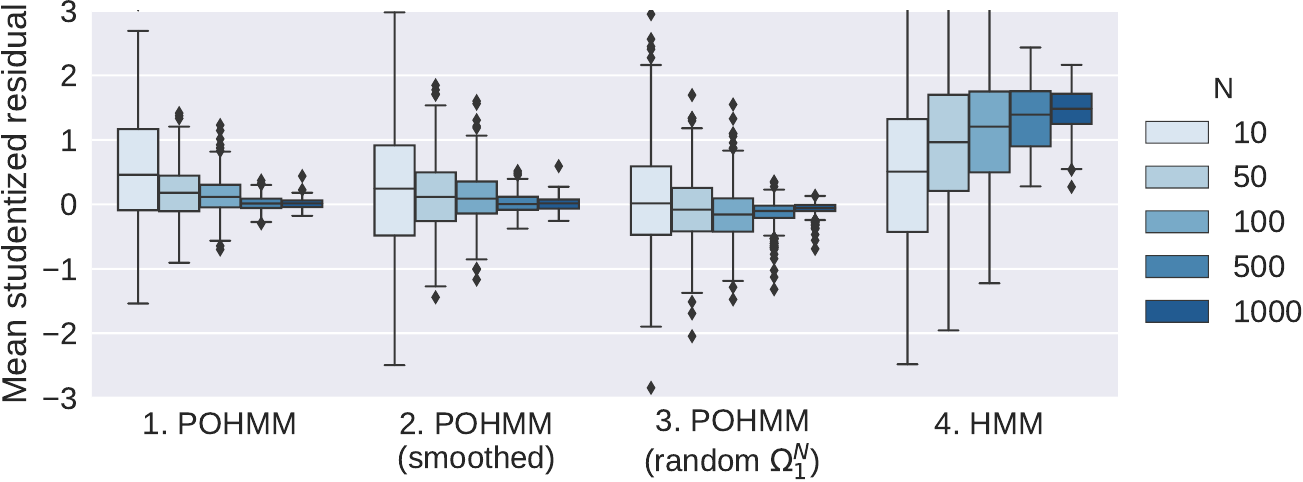} 
\par\end{centering}
}\subfloat[Hidden state classification accuracy.\label{fig:Hidden-state-classification}]{\begin{centering}
\includegraphics[width=0.3\columnwidth]{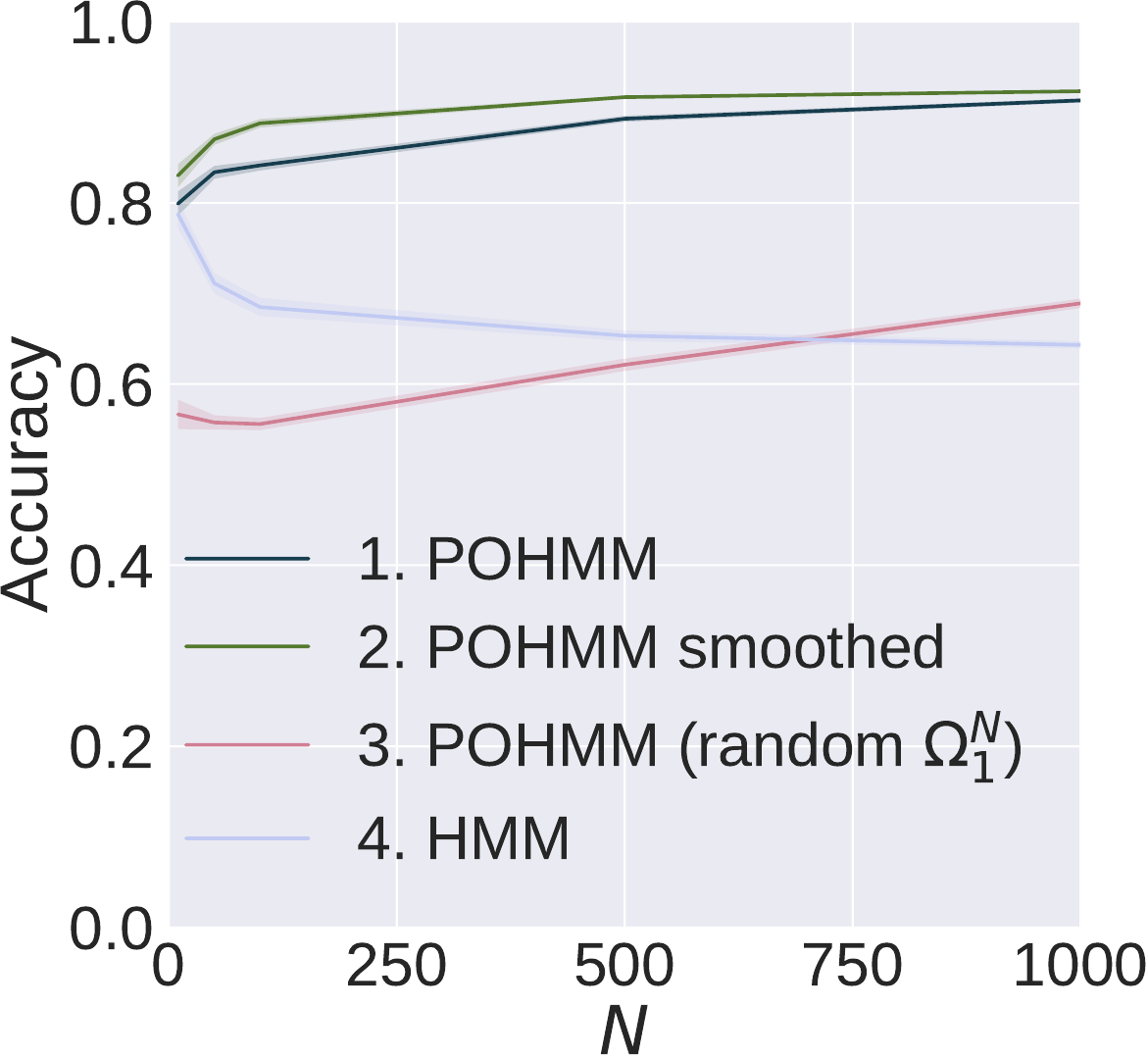} 
\par\end{centering}
}
\par\end{centering}
\caption{Simulation study results. In 1 and 2, a POHMM is trained on data generated
from a POHMM; in 3, a POHMM is trained on data generated from an HMM
(using random event types); in 4, an HMM is trained on data generated
from a POHMM (ignoring event types).}
\end{figure}

Figure \ref{fig:Studentized-residuals.} contains the mean studentized
residuals for emission parameters of each model, and Figure \ref{fig:Hidden-state-classification}
shows the hidden state classification accuracies (where chance accuracy
is $\frac{1}{2^{N}}$). Both the unsmoothed and smoothed POHMM residuals
tend toward 0 as $N$ increases, indicating consistency. The marginal
residuals for the POHMM with random event types also appear unbiased,
an indication that the POHMM marginals, in which the event type is
marginalized out, are asymptotically equivalent to the HMM. Finally,
the HMM residuals, when trained on data generated from a POHMM, appear
biased as expected when the event types are ignored. Similar results
in all scenarios are seen for the transition probability residuals
(not shown), and we confirmed that these results are insensitive to
the choice of $\theta_{o}$.

\section{Case study: keystroke dynamics\label{sec:case-study}}

Five publicly-available keystroke datasets are analyzed in this work,
summarized in Table \ref{tab:Keystroke-data-summary}. We categorize
the input type as follows:
\begin{itemize}
\item \textbf{Fixed-text}: The keystrokes exactly follow a relatively short
predefined sequence, e.g., passwords and phone numbers.
\item \textbf{Constrained-text}: The keystrokes roughly follow a predefined
sequence, e.g., case-insensitive passphrases and transcriptions. Some
massively open online course (MOOC) providers require the student
to copy several sentences for the purpose of keystroke dynamics-based
verification \citep{maas2014offering}.
\item \textbf{Free-text}: The keystrokes do not follow a predefined sequence,
e.g., responding to an open-ended question in an online exam.
\end{itemize}
The \emph{password}, \emph{keypad}, and \emph{mobile} datasets contain
short fixed-text input in which all the users in each dataset typed
the same 10-character string followed by the Enter key: ``.tie5Roanl''
for the password dataset \citep{killourhy2009comparing} and ``9141937761''
for the keypad \citep{bakelman2013keystroke} and mobile datasets
\citep{coakley2016keystroke}. Samples that contained errors or more
than 11 keystrokes were discarded. The password dataset was collected
on a laptop keyboard equipped with a high-resolution clock (estimated
resolution to within \textpm 200 \textgreek{m}s \citep{killourhy2008effect}),
while the timestamps in all other datasets were recorded with millisecond
resolution (see discussion in Section \ref{sec:Keystroke-dynamics}
on timestamp resolution). The keypad dataset used only the 10-digit
numeric keypad located on the right side a standard desktop keyboard,
and the mobile dataset used an Android touchscreen keypad with similar
layout. In addition to timestamps, the mobile dataset contains accelerometer,
gyroscope, screen location, and pressure sensor features measured
on each key press and release.

\begin{table}
\caption{Keystroke dataset summary. Columns 4-7 indicate: number of users,
samples per user, keystrokes per sample, and $\bar{\tau}$=mean press-press
latency (ms).\label{tab:Keystroke-data-summary}}
\centering{}\renewcommand{\arraystretch}{1.0}%
\begin{tabular}{|l|l|l|r|r|r|r|}
\hline 
{\scriptsize{}Dataset} & {\scriptsize{}Source} & {\scriptsize{}Category} & {\scriptsize{}Users} & {\scriptsize{}Samples/user} & {\scriptsize{}Keys/sample} & {\scriptsize{}$\bar{\tau}$ (ms)}\tabularnewline
\hline 
\hline 
{\scriptsize{}Password} & {\scriptsize{}\citep{killourhy2009comparing}} & {\scriptsize{}Short fixed} & {\scriptsize{}51} & {\scriptsize{}400} & {\scriptsize{}11} & {\scriptsize{}249}\tabularnewline
\hline 
{\scriptsize{}Keypad} & {\scriptsize{}\citep{bakelman2013keystroke}} & {\scriptsize{}Short fixed} & {\scriptsize{}30} & {\scriptsize{}20} & {\scriptsize{}11} & {\scriptsize{}376}\tabularnewline
\hline 
{\scriptsize{}Mobile} & {\scriptsize{}\citep{coakley2016keystroke}} & {\scriptsize{}Short fixed} & {\scriptsize{}51} & {\scriptsize{}20} & {\scriptsize{}11} & {\scriptsize{}366}\tabularnewline
\hline 
{\scriptsize{}Fable} & {\scriptsize{}\citep{monaco2013recent}} & {\scriptsize{}Long constrained} & {\scriptsize{}60} & {\scriptsize{}4} & {\scriptsize{}100} & {\scriptsize{}264}\tabularnewline
\hline 
{\scriptsize{}Essay} & {\scriptsize{}\citep{villani2006keystroke}} & {\scriptsize{}Long free} & {\scriptsize{}55} & {\scriptsize{}6} & {\scriptsize{}500} & {\scriptsize{}284}\tabularnewline
\hline 
\end{tabular}
\end{table}

The \emph{fable} dataset contains long constrained-text input from
60 users who each copied 4 different fables or nursery rhymes \citep{monaco2013recent,villani2006keystroke}.
Since mistakes were permitted, the keystrokes for each copy task varied,
unlike the short fixed-text datasets above. The \emph{essay} dataset
contains long free-text input from 55 users who each answered 6 essay-style
questions as part of a class exercise \citep{villani2006keystroke}.
Both the fable and essay datasets were collected on standard desktop
and laptop keyboards. For this work, the fable samples were truncated
to each contain exactly 100 keystrokes and the essay samples to each
contain exactly 500 keystrokes.

Each keystroke event contains two timing features,
\begin{eqnarray}
\tau_{n} & = & t_{n}^{P}-t_{n-1}^{P}\label{eq:press-press-latency}\\
d_{n} & = & t_{n}^{R}-t_{n}^{P}\label{eq:keystroke-duration}
\end{eqnarray}
where $t_{n}^{P}$ and $t_{n}^{R}$ are the press and release timestamps
of the $n^{th}$ keystroke, respectively; $\tau_{n}$ is the press-press
time interval and $d_{n}$ is the key-hold duration. Note that other
timing features, such as release-release and release-press intervals,
can be calculated by a linear combination of the above two features.

Each user's keystroke dynamics are modeled by a POHMM with log-normal
emission and two hidden states, all conditioned on the keyboard keys
as the observed event types. A two-state model is the simplest model
of non-homogeneous behavior, as one state implies a sequence of independent
and identically distributed (i.i.d.) observations. The two hidden
states correspond to the active and passive states of the user, in
which relatively longer time intervals are observed in the passive
state. Given the hidden state and the observed event type, the keystroke
time intervals $\tau_{n}$ and $d_{n}$ are each modeled by a log-normal
distribution (Equation \ref{eq:log-normal}), where $\eta\left[j|\omega\right]$
and $\rho\left[j|\omega\right]$ are the log-mean and log-standard
deviation, respectively, in hidden state $j$ given observed key $\omega$.

\begin{figure}[t]
\begin{centering}
\subfloat[Fixed-text]{\begin{centering}
\includegraphics[width=0.35\columnwidth]{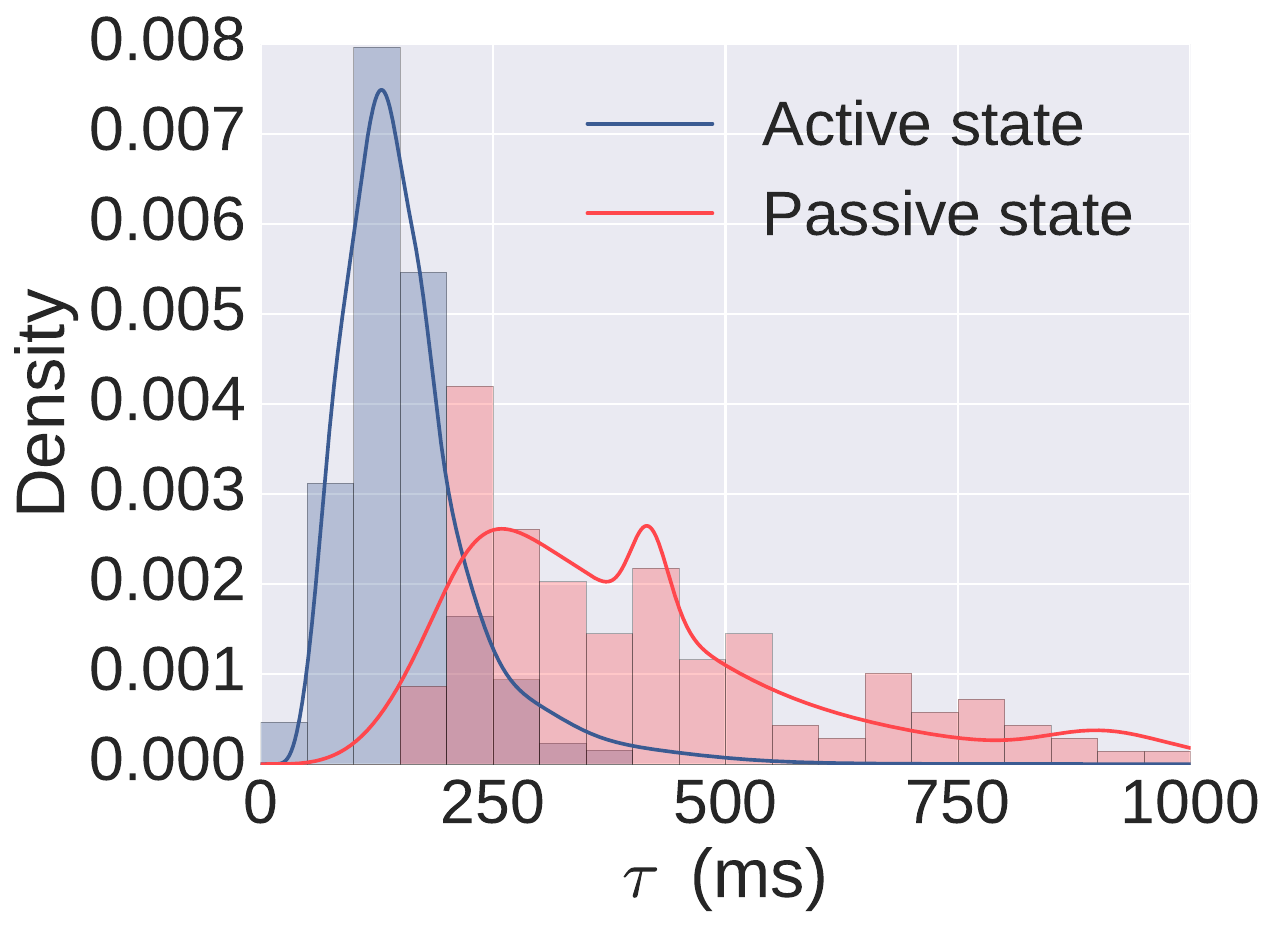}
\par\end{centering}
}\subfloat[Free-text]{\begin{centering}
\includegraphics[width=0.35\columnwidth]{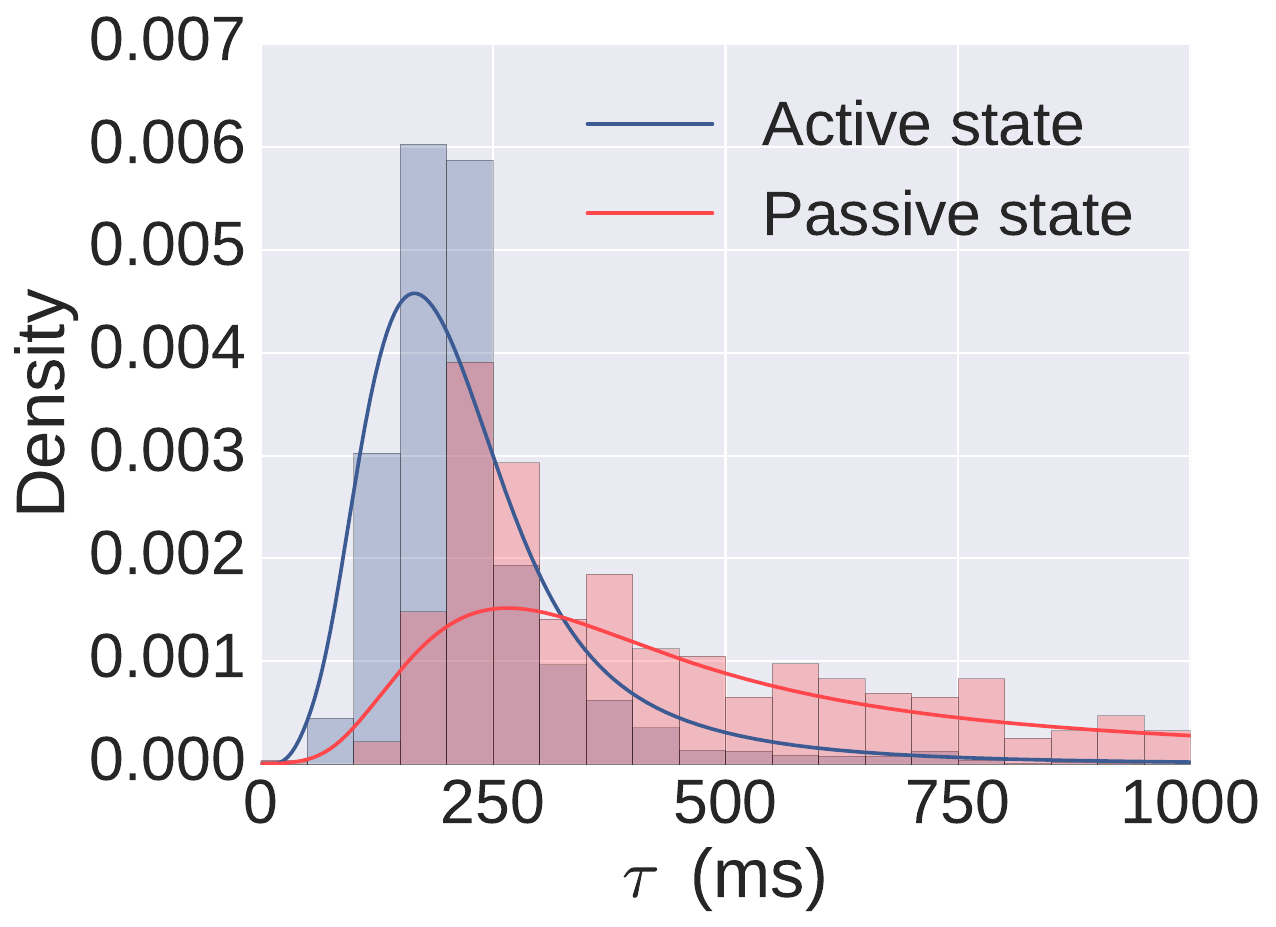}
\par\end{centering}
}
\par\end{centering}
\caption{POHMM marginal distributions showing a separation between active and
passive typing states. The marginal distributions are mixtures of
log-normals conditioned on the key names. Histograms show the empirical
time interval distributions in each hidden state. \label{fig:keystroke-pohmm-examples}}
\end{figure}

The POHMM parameters are determined using Algorithm \ref{alg:pohmm-baum-welch-algorithm-with-smoothing},
and convergence is achieved after a loglikelihood increase less than
$10^{-6}$ or 1000 iterations, whichever is reached first. As an example,
the marginal key-press time interval distributions for each hidden
state are shown in Figure \ref{fig:keystroke-pohmm-examples} for
two randomly selected samples. The passive state in the free-text
model has a heavier tail than the fixed-text, while the active state
distributions in both models are comparable. The rest of this section
presents experimental results for a goodness of fit test, identification,
verification, and continuous verification. Source code to reproduce
the experiments in this article is available\footnote{Code to reproduce experiments: \href{https://github.com/vmonaco/pohmm-keystroke}{https://github.com/vmonaco/pohmm-keystroke}}.

\subsection{Goodness of fit}

To determine whether the POHMM is consistent with observed data, a
Monte Carlo goodness of fit test is performed. The test proceeds as
follows. For each keystroke sample (using the key-press time intervals
only), the model parameters $\hat{\theta}_{m}$ are determined. The
area test statistic between the model and empirical distribution is
then taken. The area test statistic is a compromise between the Kolmogorov-Smirnov
(KS) test and Cramér-von Mises test \citep{malmgren2008poissonian},
\begin{equation}
A=\int|P_{D}(\tau)-P_{M}\left(\tau|\hat{\theta}_{m}\right)|\mathrm{d}\tau\label{eq:area-test-stat}
\end{equation}
where $P_{D}$ is the empirical cumulative distribution and $P_{M}$
is the model cumulative distribution. The POHMM marginal emission
density is given by
\begin{equation}
g\left({\bf x};\theta\right)=\sum_{\omega\in\Omega}\sum_{j=1}^{M}\Pi\left[\omega\right]\Pi\left[j\right]f\left({\bf x};{\bf b}\left[j|\omega\right]\right)\label{eq:marginal-dist}
\end{equation}
where $\Pi\left[j\right]$ is the stationary probability of hidden
state $j$ and $\Pi\left[\omega\right]$ is the stationary probability
of event type $\omega$. Using the fitted model with parameters $\hat{\theta}_{m}$,
a surrogate data sample the same size as the empirical sample is generated.
Estimated parameters $\hat{\theta}_{s}$ are determined using the
surrogate sample in a similar fashion as the empirical sample. The
area test statistic between the surrogate-data-trained model and surrogate
data is computed, given by $A_{s}$. This process repeats until enough
surrogate statistics have accumulated to reliably determine $P(|A_{s}-\langle A_{s}\rangle|>|A-\langle A_{s}\rangle|)$.
The biased p-value is given by
\begin{equation}
\frac{I\left(|A_{s}-\langle A_{s}\rangle|>|A-\langle A_{s}\rangle|\right)+1}{S+1}\label{eq:p-value}
\end{equation}
where $I\left(\cdot\right)$ is the indicator function. Testing the
null hypothesis, that the model is consistent with the data, requires
fitting $S+1$ models (1 empirical and $S$ surrogate samples).

\begin{figure}[t]
\begin{centering}
\subfloat[Constrained-text]{\begin{centering}
\includegraphics[width=0.35\columnwidth]{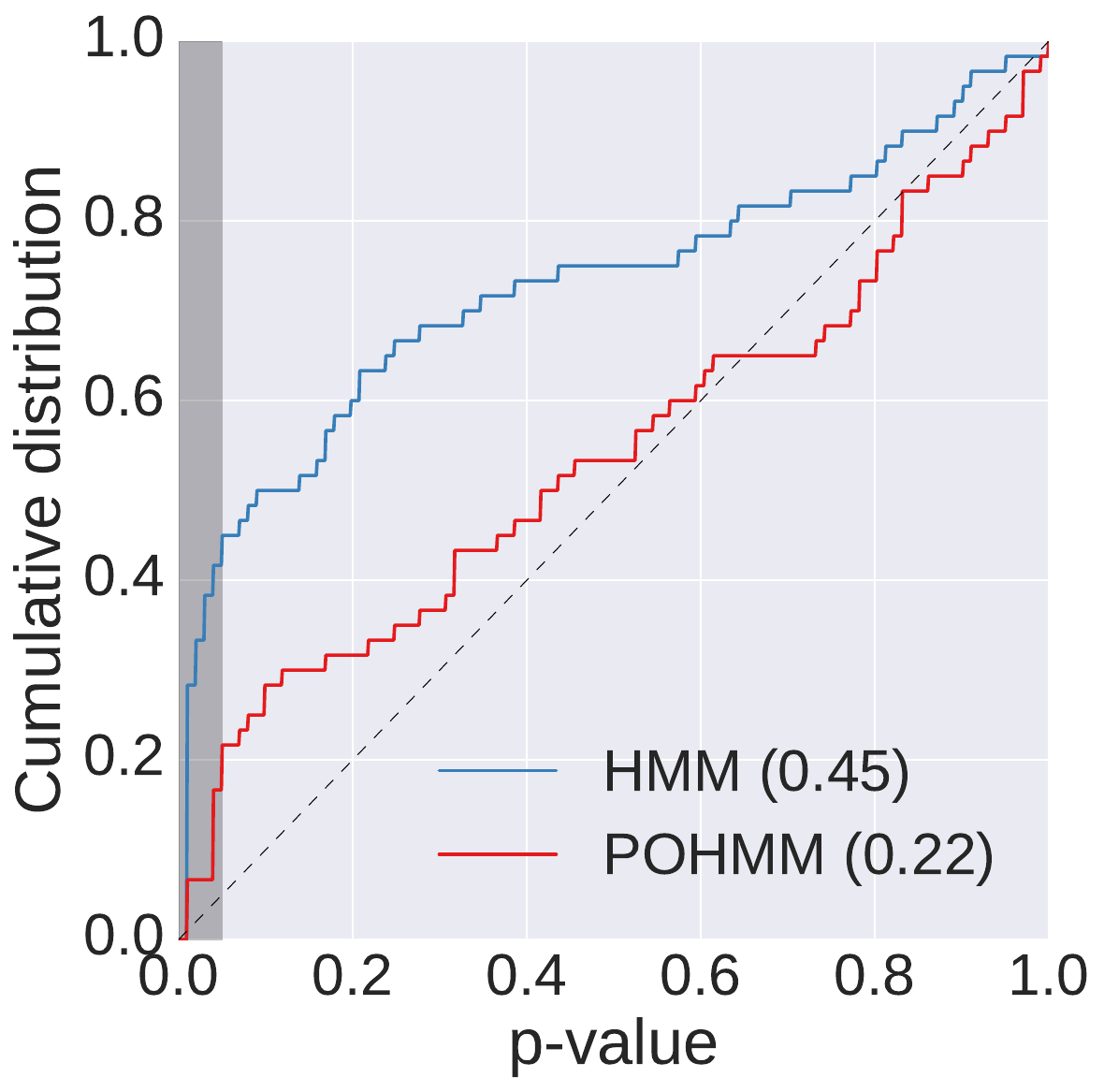}
\par\end{centering}
}\subfloat[Free-text]{\begin{centering}
\includegraphics[width=0.35\columnwidth]{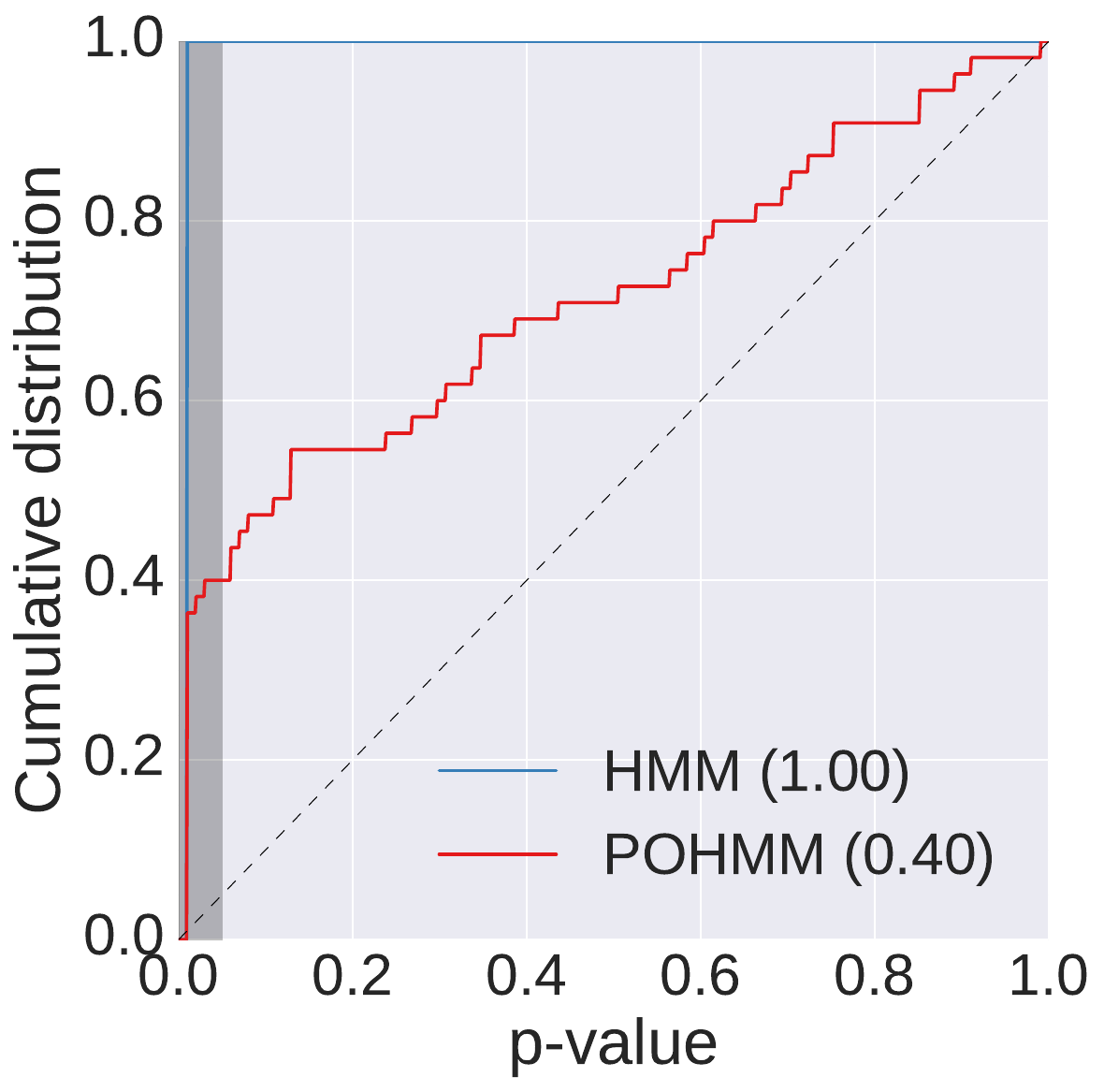}
\par\end{centering}
}
\par\end{centering}
\centering{}\caption{Keystroke goodness of fit p-value distributions testing the null hypothesis
that the model is consistent with the data. Proportions of rejected
samples at the 0.05 significance level are shown in parentheses. If
the null hypothesis was true, i.e., the model was actually consistent
with the keystroke data, then p-values would follow a uniform distribution
shown by the dashed black line.\label{fig:keystroke-p-values}}
\end{figure}

The test is performed for both the HMM and the POHMM for each user
in the fable and essay datasets, using the key-press time intervals
only. The resulting p-value distributions are shown in Figure \ref{fig:keystroke-p-values}.
The shaded area represents a 0.05 significance level in which the
null hypothesis is rejected. In the fable dataset, the HMM is rejected
for 45\% of users, while the POHMM is rejected for 22\% of users.
The HMM is rejected for 100\% of users in the essay dataset, and the
POHMM is rejected for 40\% of users. If the POHMM truly reflected
typing behavior (i.e., the null hypothesis was actually true), the
p-values would follow a uniform distribution shown by the dashed black
line. In both experiments, the POHMM is largely preferred over the
HMM.

\subsection{Identification and verification}

We use the POHMM to perform both user identification and verification,
and compare the results to other leading methods. Identification,
a multiclass classification problem, is performed by the MAP approach
in which the model with maximum a posterior probability is chosen
as the class label. This approach is typical in using a generative
model to perform classification. Better performance could, perhaps,
be achieved through parameter estimation with a discriminative criterion
\citep{MUTSAM201614}, or a hybrid discriminative/generative model
in which the POHMM parameters provide features for a discriminative
classifier \citep{BICEGO20092637}. Verification, a binary classification
problem, is achieved by comparing the claimed user's model loglikelihood
to a threshold.

Identification and verification results are obtained for each keystroke
dataset and four benchmark anomaly detectors in addition to the POHMM.
The password dataset uses a validation procedure similar to \citet{killourhy2009comparing},
except only samples from the 4th session (repetitions 150-200) are
used for training and sessions 5-8 (repetitions 201-400) for testing.
For the other datasets, results are obtained through a stratified
cross-fold validation procedure with the number of folds equal to
the number of samples per user: 20 for keypad and mobile, 4 for fable,
and 6 for essay. In each fold, one sample from each user is retained
as a query and the remaining samples are used for training.

Identification accuracy (ACC) is measured by the proportion of correctly
classified query samples. Verification performance is measured by
the user-dependent equal error rate (EER), the point on the receiver
operating characteristic (ROC) curve at which the false rejection
rate (FRR) and false acceptance rate (FAR) are equal. Each query sample
is compared against every model in the population, only one of which
will be genuine. The resulting loglikelihood is normalized using the
minimum and maximum loglikelihoods from every model in the population
to obtain a normalized score between 0 and 1. Confidence intervals
for both the ACC and EER are obtained over users in each dataset,
similar to \citep{killourhy2009comparing}.

\begin{table}
\caption{Identification accuracy rates. Bold indicates systems that are not
significantly worse than the best system. Mobile+ includes mobile
sensor features in addition to time intervals.\label{tab:identification-results}}
\centering{}\renewcommand{\arraystretch}{1.0}%
\begin{tabular}{|l|r|r|r|r|r|}
\cline{2-6} 
\multicolumn{1}{l|}{} & {\scriptsize{}Manhattan} & {\scriptsize{}Manhattan (Scaled)} & {\scriptsize{}SVM (One-class)} & {\scriptsize{}HMM} & {\scriptsize{}POHMM}\tabularnewline
\hline 
{\scriptsize{}Password} & {\scriptsize{}0.510 (0.307)} & {\scriptsize{}0.662 (0.282)} & {\scriptsize{}0.465 (0.293)} & {\scriptsize{}0.467 (0.295)} & \textbf{\scriptsize{}0.789 (0.209)}\tabularnewline
\hline 
{\scriptsize{}Keypad} & \textbf{\scriptsize{}0.623 (0.256)} & \textbf{\scriptsize{}0.713 (0.200)} & {\scriptsize{}0.500 (0.293)} & {\scriptsize{}0.478 (0.287)} & \textbf{\scriptsize{}0.748 (0.151)}\tabularnewline
\hline 
{\scriptsize{}Mobile} & {\scriptsize{}0.290 (0.230)} & {\scriptsize{}0.528 (0.237)} & {\scriptsize{}0.267 (0.229)} & {\scriptsize{}0.303 (0.265)} & \textbf{\scriptsize{}0.607 (0.189)}\tabularnewline
\hline 
{\scriptsize{}Mobile+} & {\scriptsize{}0.647 (0.250)} & \textbf{\scriptsize{}0.947 (0.104)} & {\scriptsize{}0.857 (0.232)} & {\scriptsize{}0.937 (0.085)} & \textbf{\scriptsize{}0.971 (0.039)}\tabularnewline
\hline 
{\scriptsize{}Fable} & {\scriptsize{}0.492 (0.332)} & {\scriptsize{}0.613 (0.314)} & {\scriptsize{}0.571 (0.235)} & {\scriptsize{}0.392 (0.355)} & \textbf{\scriptsize{}0.887 (0.175)}\tabularnewline
\hline 
{\scriptsize{}Essay} & {\scriptsize{}0.730 (0.320)} & \textbf{\scriptsize{}0.839 (0.242)} & {\scriptsize{}0.342 (0.302)} & {\scriptsize{}0.303 (0.351)} & \textbf{\scriptsize{}0.909 (0.128)}\tabularnewline
\hline 
\end{tabular}
\end{table}

Benchmark anomaly detectors include Manhattan distance, scaled Manhattan
distance, one-class support vector machine (SVM), and a two-state
HMM. The Manhattan, scaled Manhattan, and one-class SVM operate on
fixed-length feature vectors, unlike the HMM and POHMM. Timing feature
vectors for the password, keypad, and mobile datasets are formed by
the 11 press-press latencies and 10 durations of each 11-keystroke
sample for a total of 21 timing features. The mobile sensors provide
an additional 10 features for each keystroke event for a total of
131 features. For each event, the sensor features include: acceleration
(meters/second\textsuperscript{2}) and rotation (radians/second)
along three orthogonal axes (6 features), screen coordinates (2 features),
pressure (1 feature), and the length of the major axis of an ellipse
fit to the pointing device (1 feature). Feature vectors for the fable
and essay datasets are each comprised of a set of 218 descriptive
statistics for various keystroke timings. Such timing features include
the sample mean and standard deviation of various sets of key durations,
e.g., consonants, and latency between sets of keys, e.g., from consonants
to vowels. For a complete list of features see \citep{monaco2013recent,tappert2009keystroke}.
The feature extraction also includes a rigorous outlier removal step
that excludes observations outside a specified confidence interval
and a hierarchical fallback scheme that accounts for missing or infrequent
observations.

The Manhattan anomaly detector uses the negative Manhattan distance
to the mean template vector as a confidence score. For the scaled
Manhattan detector, features are first scaled by the mean absolute
deviation over the entire dataset. This differs slightly from the
scaled Manhattan in \citep{killourhy2009comparing}, which uses the
mean absolute deviation of each user template. The global (over the
entire dataset) mean absolute deviation is used in this work due to
the low number of samples per user in some datasets. The one-class
SVM uses a radial basis function (RBF) kernel and 0.5 tolerance of
training errors, i.e., half the samples will become support vectors.
The HMM is exactly the same as the POHMM (two hidden states and log-normal
emissions), except event types are ignored.

\begin{figure}[t]
\begin{centering}
\subfloat[Password]{\begin{centering}
\includegraphics[width=0.25\columnwidth]{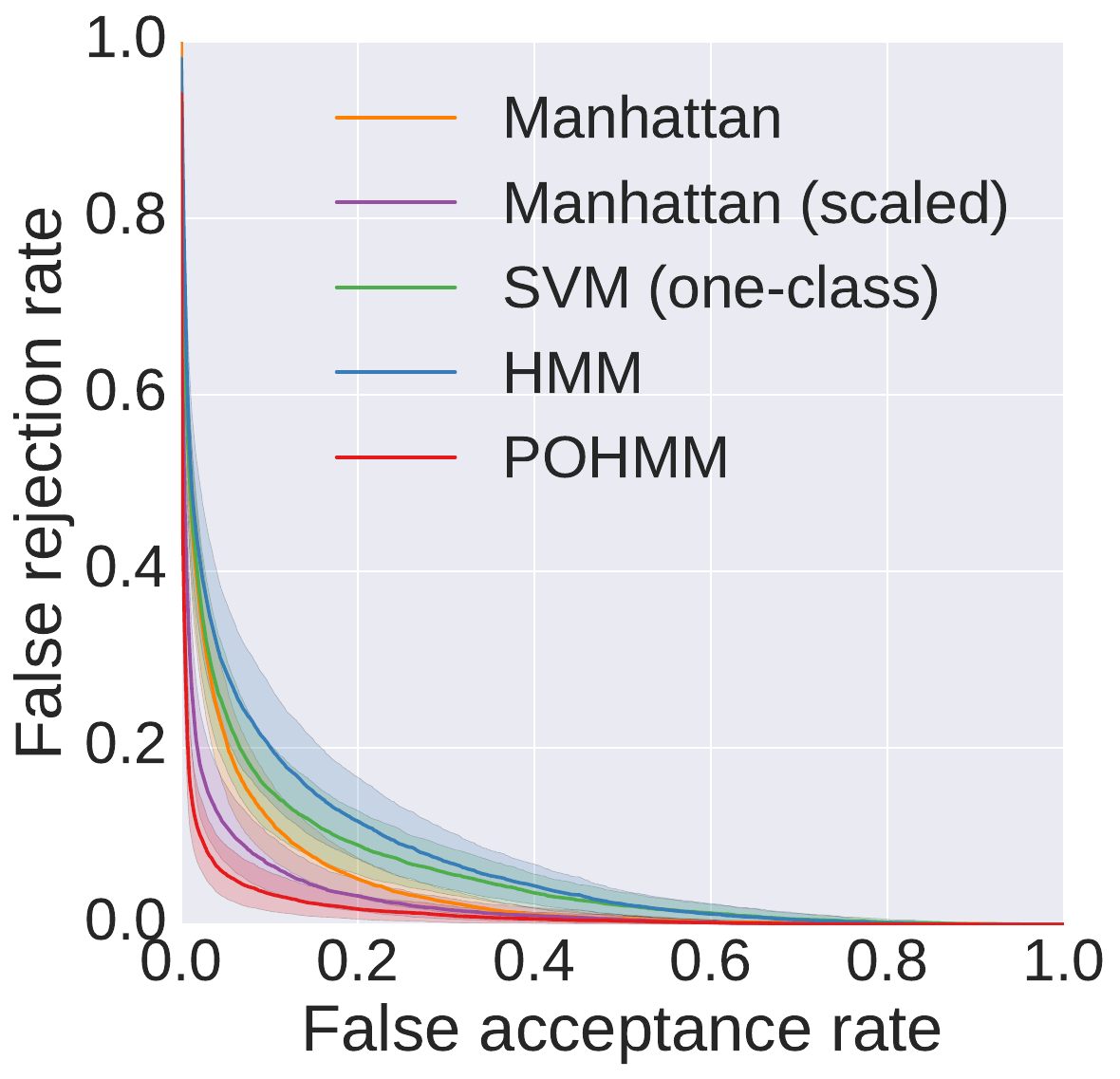}
\par\end{centering}
}\subfloat[Keypad]{\begin{centering}
\includegraphics[width=0.25\columnwidth]{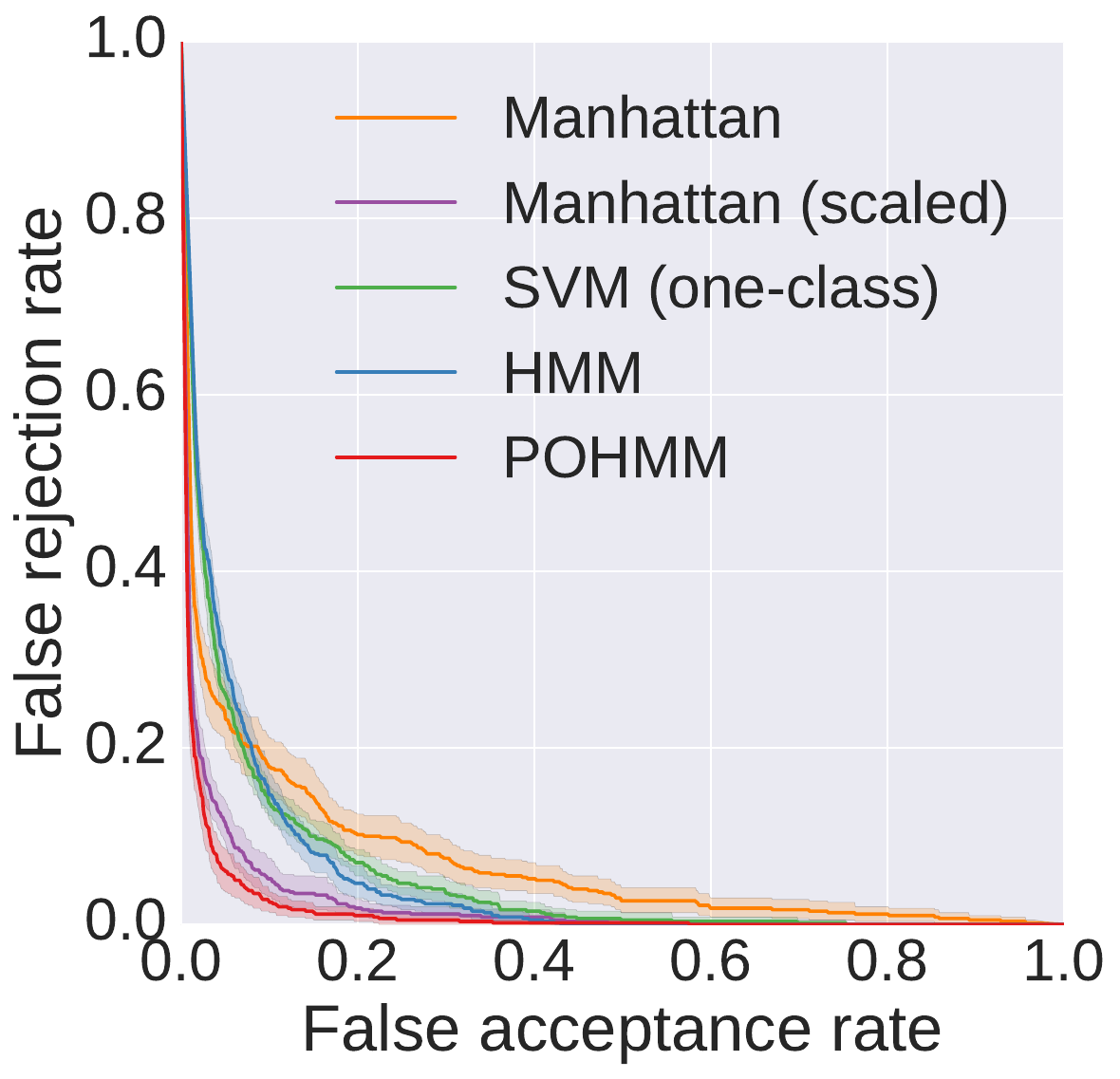}
\par\end{centering}
}\subfloat[Fixed-text]{\begin{centering}
\includegraphics[width=0.25\columnwidth]{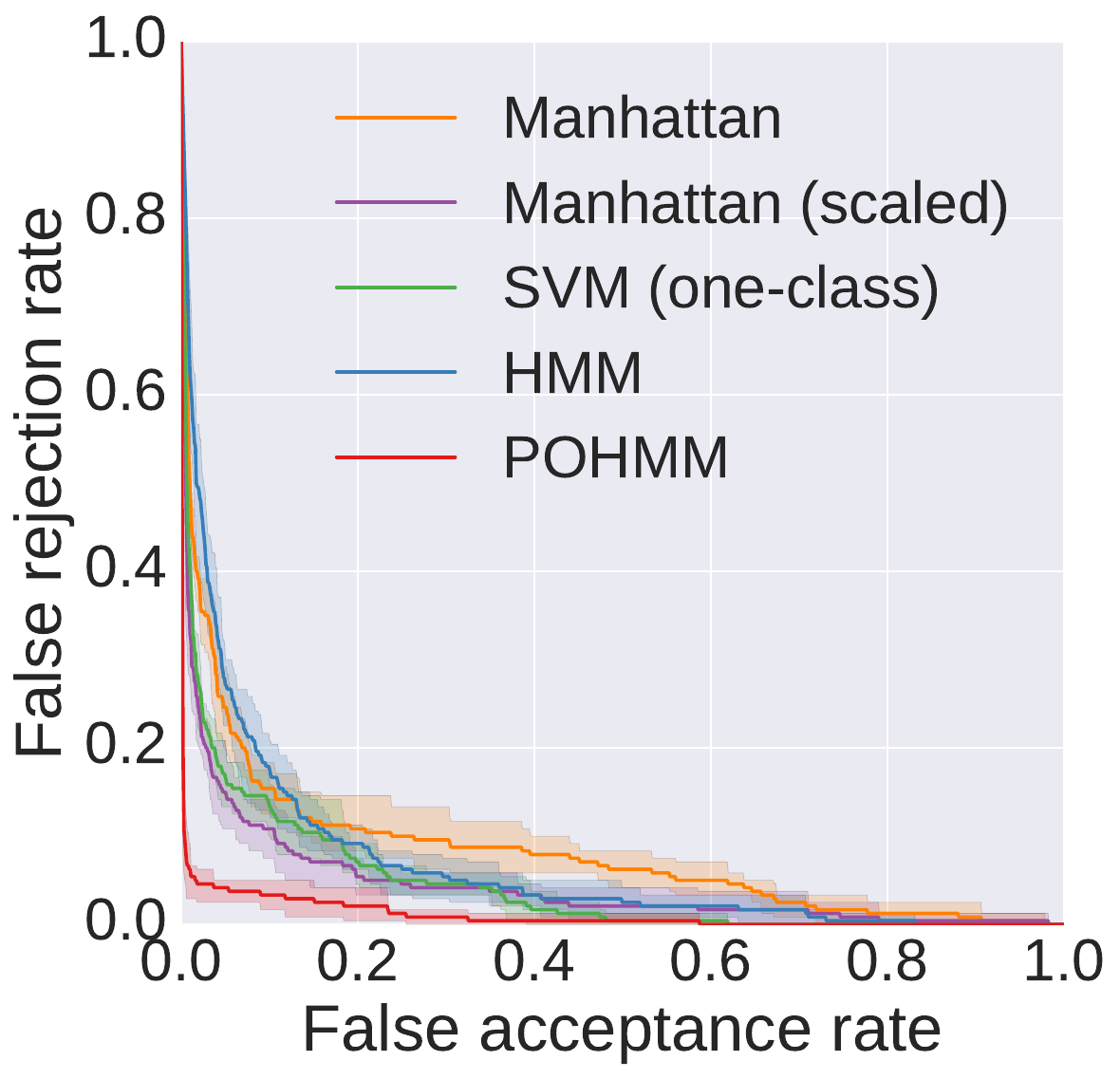}
\par\end{centering}
}\subfloat[Free-text]{\begin{centering}
\includegraphics[width=0.25\columnwidth]{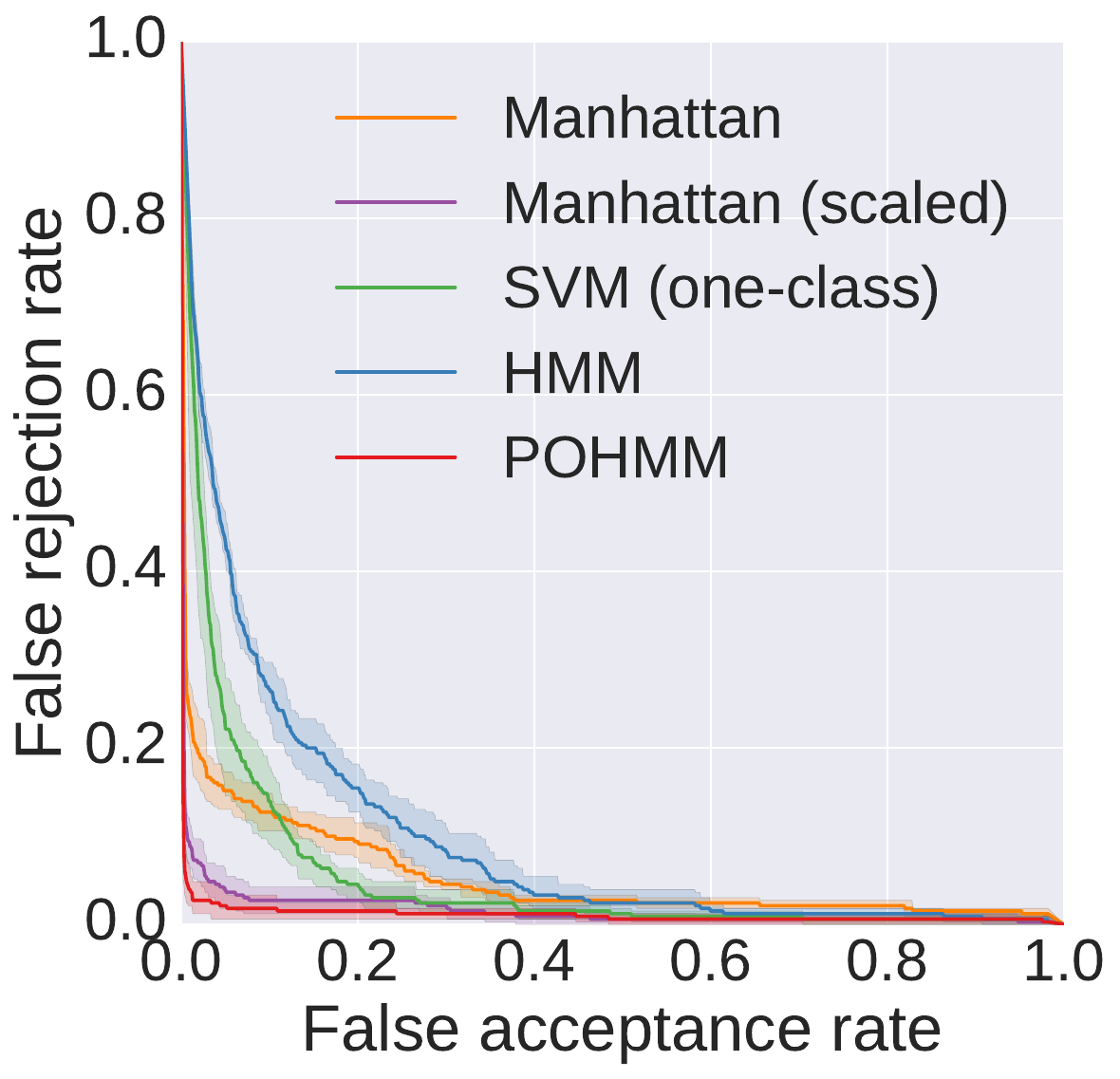}
\par\end{centering}
}
\par\end{centering}
\caption{Keystroke ROC curves. Bands show the 95\% confidence intervals.\label{fig:Keystrokes-ROC-curves}}
\end{figure}

\begin{table}
\caption{User-dependent EER. Bold indicates systems that are not significantly
worse than the best system. Mobile+ includes mobile sensor features
in addition to time intervals.\label{tab:verification-results}}
\centering{}\renewcommand{\arraystretch}{1.0}%
\begin{tabular}{|l|r|r|r|r|r|}
\cline{2-6} 
\multicolumn{1}{l|}{} & {\scriptsize{}Manhattan} & {\scriptsize{}Manhattan (scaled)} & {\scriptsize{}SVM (one-class)} & {\scriptsize{}HMM} & {\scriptsize{}POHMM}\tabularnewline
\hline 
{\scriptsize{}Password} & {\scriptsize{}0.088 (0.069)} & {\scriptsize{}0.062 (0.064)} & {\scriptsize{}0.112 (0.088)} & {\scriptsize{}0.126 (0.099)} & \textbf{\scriptsize{}0.042 (0.051)}\tabularnewline
\hline 
{\scriptsize{}Keypad} & {\scriptsize{}0.092 (0.069)} & \textbf{\scriptsize{}0.053 (0.030)} & {\scriptsize{}0.110 (0.054)} & {\scriptsize{}0.099 (0.050)} & \textbf{\scriptsize{}0.053 (0.025)}\tabularnewline
\hline 
{\scriptsize{}Mobile} & {\scriptsize{}0.194 (0.101)} & \textbf{\scriptsize{}0.097 (0.057)} & {\scriptsize{}0.170 (0.092)} & {\scriptsize{}0.168 (0.085)} & \textbf{\scriptsize{}0.090 (0.054)}\tabularnewline
\hline 
{\scriptsize{}Mobile+} & {\scriptsize{}0.084 (0.061)} & \textbf{\scriptsize{}0.009 (0.027)} & \textbf{\scriptsize{}0.014 (0.033)} & {\scriptsize{}0.013 (0.021)} & \textbf{\scriptsize{}0.006 (0.014)}\tabularnewline
\hline 
{\scriptsize{}Fable} & {\scriptsize{}0.085 (0.091)} & {\scriptsize{}0.049 (0.060)} & {\scriptsize{}0.099 (0.106)} & {\scriptsize{}0.105 (0.092)} & \textbf{\scriptsize{}0.031 (0.077)}\tabularnewline
\hline 
{\scriptsize{}Essay} & {\scriptsize{}0.061 (0.092)} & {\scriptsize{}0.028 (0.052)} & {\scriptsize{}0.098 (0.091)} & {\scriptsize{}0.145 (0.107)} & \textbf{\scriptsize{}0.020 (0.046)}\tabularnewline
\hline 
\end{tabular}
\end{table}

Identification and verification results are shown in Tables \ref{tab:identification-results}
and \ref{tab:verification-results}, respectively, and ROC curves
are shown in Figure \ref{fig:Keystrokes-ROC-curves}. The best-performing
anomaly detectors in Tables \ref{tab:identification-results} and
\ref{tab:verification-results} are shown in bold. The set of best-performing
detectors contains those that are not significantly worse than the
POHMM, which achieves the highest performance in every experiment.
The Wilcoxon signed-rank test is used to determine whether a detector
is significantly worse than the best detector, testing the null hypothesis
that a detector has the same performance as the POHMM. A Bonferroni
correction is applied to control the family-wise error rate, i.e.,
the probability of falsely rejecting a detector that is actually in
the set of best-performing detectors \citep{rupert2012simultaneous}.
At a 0.05 significance level, the null hypothesis is rejected with
a p-value not greater than $\frac{0.05}{4}$ since four tests are
applied in each row. The POHMM achieves the highest identification
accuracy and lowest equal error rate for each dataset. For 3 out of
6 datasets in both sets of experiments, all other detectors are found
to be significantly worse than the POHMM.

\subsection{Continuous verification}

Continuous verification has been recognized as a problem in biometrics
whereby a resource is continuously monitored to detect the presence
of a genuine user or impostor \citep{sim2007continuous}. It is natural
to consider the continuous verification of keystroke dynamics, and
most behavioral biometrics, since events are continuously generated
as the user interacts with the system. In this case, it is desirable
to detect an impostor within as few keystrokes as possible. This differs
from the static verification scenario in the previous section in which
verification performance is evaluated over an entire session. Instead,
continuous verification requires a verification decision to be made
upon each new keystroke \citep{bours2015performance}.

Continuous verification is enforced through a penalty function in
which each new keystroke incurs a non-negative penalty within a sliding
window. The penalty at any given time can be thought of as the inverse
of trust. As behavior becomes more consistent with the model, the
cumulative penalty within the window can decrease, and as it becomes
more dissimilar, the penalty increases. The user is rejected if the
cumulative penalty within the sliding window exceeds a threshold.
The threshold is chosen for each sample such that the genuine user
is never rejected, analogous to a 0\% FRR in static verification.
An alternative to the penalty function is the penalty-and-reward function
in which keystrokes incur either a penalty or a reward (i.e., a negative
penalty) \citep{bours2012continuous}. In this work, the sliding window
replaces the reward since penalties outside the window do not contribute
towards the cumulative penalty.

The penalty of each new event is determined as follows. The marginal
probability of each new event, given the preceding events, is obtained
from the forward lattice, $\alpha$, given by
\begin{equation}
P\left({\bf x}_{n+1}|{\bf x}_{1}^{n}\right)=P\left({\bf x}_{1}^{n+1}\right)-P\left({\bf x}_{1}^{n}\right)\label{eq:event-marginal-proba}
\end{equation}
When a new event is observed, the likelihood is obtained under every
model in a population of $U$ models. The likelihoods are ranked,
with the highest model given a rank of 0, and the lowest a rank of
$U-1$. The rank of the claimed user's model is the incurred penalty.
Thus, if a single event is correctly matched to the genuine user's
model, a penalty of 0 is incurred; if it scores the second highest
likelihood, a penalty of 1 is incurred, etc. The rank penalty is added
to the cumulative penalty in the sliding window, while penalties outside
the window are discarded. A window of length 25 is used in this work.

Continuous verification performance is reported as the number of events
(up to the sample length) that can occur before an impostor is detected.
This is determined by increasing the penalty threshold until the genuine
user is never rejected by the system. Since the genuine user's penalty
is always below the threshold, this is the maximum number of events
that an impostor can execute before being rejected by the system while
the genuine user is never rejected.

\begin{figure}
\begin{centering}
\includegraphics[width=0.5\columnwidth]{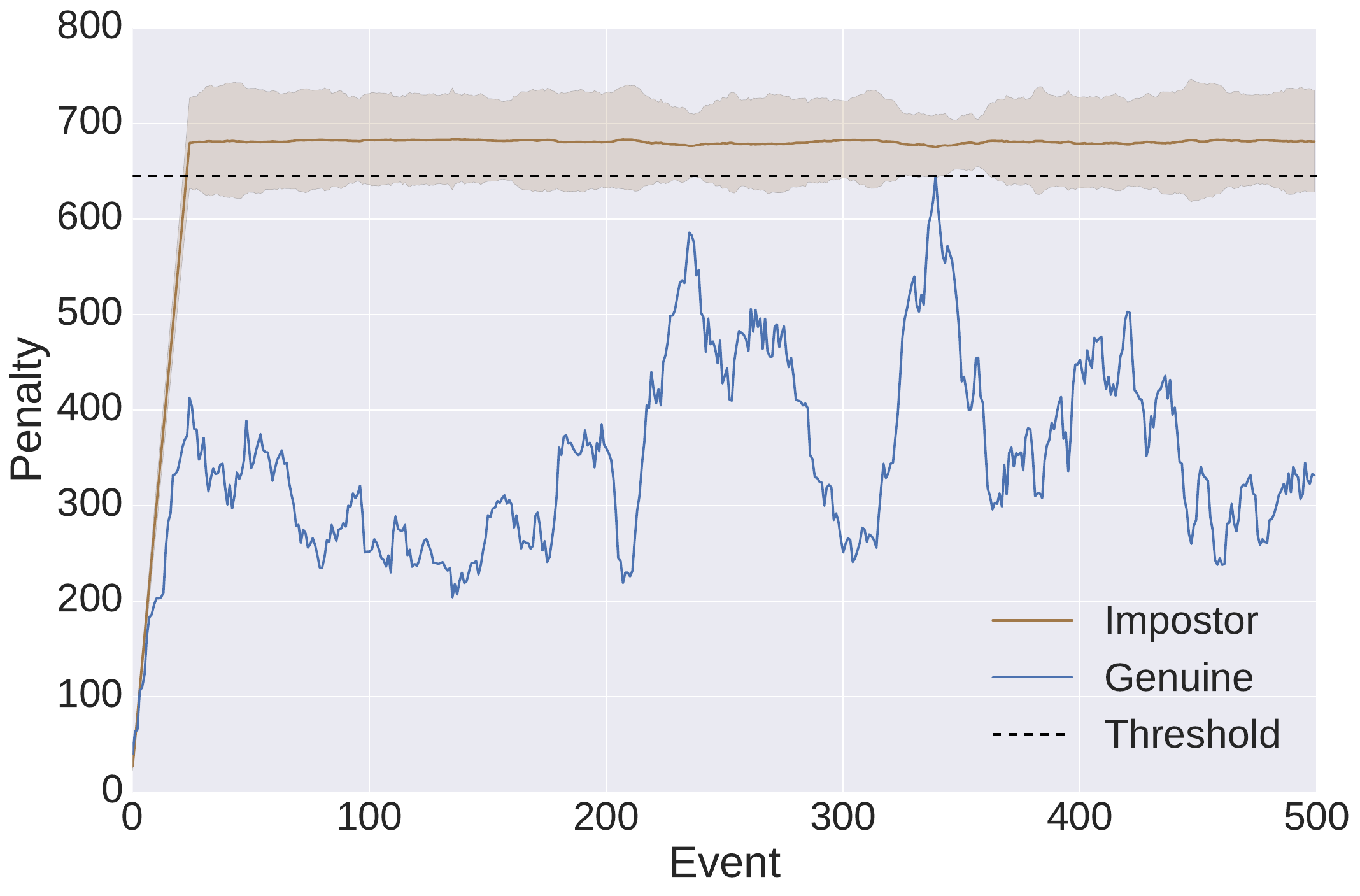}
\par\end{centering}
\caption{Continuous verification example. Bands show the 95\% confidence interval.
In this example, impostors are detected after an average of 81 keystrokes.\label{fig:Penalty-function-example}}
\end{figure}

An example of the penalty function for genuine and impostor users
is shown in Figure \ref{fig:Penalty-function-example}. The decision
threshold is set to the maximum penalty incurred by the genuine user
so that a false rejection does not occur. The average penalty for
impostor users with 95\% confidence interval is shown. In this example,
the impostor penalties exceed the decision threshold after 81 keystrokes
on average. Note that this is different than the average imposter
penalty, which exceeds the threshold after 23 keystrokes.

For each dataset, the average maximum rejection time (AMRT) is determined,
shown in Table \ref{tab:continuous-verification-results}. The maximum
rejection time (MRT) is the maximum number of keystrokes needed to
detect an impostor without rejecting the genuine user, or the time
to correct reject (TCR) with perfect usability \citep{sim2007continuous}.
The MRT is determined for each combination of impostor query sample
and user model in the dataset to get the AMRT. The POHMM has a lower
AMRT than the HMM for every dataset, and less than half that of the
HMM for free-text input.

\begin{table}[t]
\caption{Continuous verification average maximum rejection time: the number
of events that occur before an impostor is detected given the genuine
user is not falsely rejected.\label{tab:continuous-verification-results}}
\centering{}\renewcommand{\arraystretch}{1.0}%
\begin{tabular}{|l|r|r|}
\cline{2-3} 
\multicolumn{1}{l|}{} & {\scriptsize{}HMM} & {\scriptsize{}POHMM}\tabularnewline
\hline 
{\scriptsize{}Password} & {\scriptsize{}5.64 (2.04)} & {\scriptsize{}3.42 (2.04)}\tabularnewline
\hline 
{\scriptsize{}Keypad} & {\scriptsize{}4.54 (2.09)} & {\scriptsize{}3.45 (1.73)}\tabularnewline
\hline 
{\scriptsize{}Mobile} & {\scriptsize{}5.63 (2.18)} & {\scriptsize{}4.29 (2.02)}\tabularnewline
\hline 
{\scriptsize{}Mobile+} & {\scriptsize{}0.15 (0.65)} & {\scriptsize{}0.12 (0.57)}\tabularnewline
\hline 
{\scriptsize{}Fable} & {\scriptsize{}33.63 (15.47)} & {\scriptsize{}20.81 (9.07)}\tabularnewline
\hline 
{\scriptsize{}Essay} & {\scriptsize{}129.36 (95.45)} & {\scriptsize{}55.18 (68.31)}\tabularnewline
\hline 
\end{tabular}
\end{table}

\section{Discussion\label{sec:discussion}}

There have been several generalizations of the standard HMM to deal
with hidden states that are partially observable in some way. These
models are referred to as partly-HMM \citep{kobayashi1997partly},
partially-HMM \citep{ozkan2014novel}, and context-HMM \citep{forchhammer1996partially}.

The partly-HMM is a second order model in which the first state is
hidden and the second state is observable \citep{kobayashi1997partly}.
In the partly-HMM, both the hidden state and emission at time $t_{n}$
depend on the observation at time $t_{n-1}$. The partly-HMM can be
applied to problems that have a transient underlying process, such
as gesture and speech recognition, as opposed to a piecewise stationary
process that the HMM assumes \citep{iobayashi1999partly}. Parameter
estimation is performed by the EM algorithm, similar to the HMM.

Partially observable states can also come in the form of partial and
uncertain ground truth regarding the hidden state at each time step.
The partially-HMM addresses this scenario, in which an uncertain hidden
state label may be observed at each time step \citep{ozkan2014novel}.
The probability of observing the uncertain label and the probability
of the label being correct, were the true hidden state known, are
controlled by parameters $p_{obs}$ and $p_{true}$, respectively.
Thus, the probability of observing a correct label is $p_{obs}\times p_{true}$.
This model is motivated by language modeling applications in which
manually labeling data is expensive and time consuming. Similar to
the HMM, the EM algorithm can be used for estimating the parameters
of the partially-HMM \citep{ozkan2014novel}. 

Past observations can also provide context for the emission and hidden
state transition probabilities in an HMM. \citet{forchhammer1996partially}
proposed the context-HMM, in which the emission and hidden state probabilities
at time $t_{n+1}$ are conditioned on contexts $r_{n}$ and $s_{n}$,
respectively. Each context is given by a function of the previous
observations up to time $t_{n}$. The context-HMM has information
theoretic motivations, with applications such as image compression
\citep{forchhammer1999adaptive}. Used in this way, the neighboring
pixels in an image can provide context for the emission and transition
probabilities. 

There are two scenarios in which previous models of partial observability
fall short. The first is when there is missing data during parameter
estimation, such missing context, and the second is when there is
missing or novel data during likelihood calculation. A possible solution
to these problems uses the explicit marginal emission and transition
distributions, where, e.g., the context is marginalized out. While
none of the above models possess this property, the POHMM, described
in Section \ref{sec:pohmm}, has explicit marginal distributions that
are used when missing or novel data are encountered. Additionally,
parameter smoothing uses the marginal distributions to regularize
the model and improve parameter estimates.

The POHMM is different from the partly-HMM \citep{kobayashi1997partly},
being a first order model, and different from the partially-HMM \citep{ozkan2014novel},
since it doesn't assume a partial labeling. The POHMM is most similar
to the context-HMM \citep{forchhammer1996partially} in the sense
that emission and transition probabilities are conditioned on some
observed values. Despite this, there are several important differences
between the POHMM and context-HMM:
\begin{enumerate}
\item The context is not a function of the previous emissions; instead it
is a separate observed value (called an \emph{event type} in this
work).
\item The context for hidden state and emission is the same, i.e., $s_{n}=r_{n}$.
\item The emission at time $n+1$ is conditioned on a context observed at
time $n+1$ instead of time $n$.
\item An additional context $s_{n+1}$ is available at time $n+1$, upon
which the hidden state is also conditioned.
\end{enumerate}
The first difference enables the POHMM to characterize system behavior
that depends on an independent Markov chain which emanates from a
completely separate process. Such a scenario is encountered in keystroke
dynamics, whereby typing behavior depends on the text that is being
typed, but the text itself is not considered part of the keystroke
dynamics. This distinction is not made in the context-HMM, as the
context is based on the previously-observed emissions. Additionally,
the context-HMM, as original described, contains only discrete distributions
and lacks explicit marginal distributions; therefore it is unable
to account for missing or novel data during likelihood calculation,
as would be needed in free-text keystroke dynamics.

\section{Conclusions\label{sec:Conclusion}}

This work introduced the POHMM, an extension of the HMM in which the
hidden states are partially observable through an independent Markov
chain. Computational complexities of POHMM parameter estimation and
likelihood calculation are comparable to that of the HMM, which are
linear in the number of observations. POHMM parameter estimation also
inherits the desirable properties of expectation maximization, as
a modified Baum-Welch algorithm is employed. A case study of the POHMM
applied to keystroke dynamics demonstrates superiority over leading
alternative models on a variety of tasks, including identification,
verification, and continuous verification.

Since we assumed the event type is given, we considered only the conditional
likelihood $P\left({\bf x}_{1}^{N}|\Omega_{1}^{N}\right)$. Consideration
of the joint likelihood $P\left({\bf x}_{1}^{N},\Omega_{1}^{N}\right)$
remains an item for future work. Applied to keystroke dynamics, the
joint likelihood $P\left({\bf x}_{1}^{N},\Omega_{1}^{N}\right)$ would
reflect both the keystroke timings and keys typed enabling the model
to capture both typing behavior and text generation. Alternatively,
the consideration of $P\left(\Omega_{1}^{N}|{\bf x}_{1}^{N}\right)$
would enable the POHMM to recover the key names from keystroke timings,
also an item for future work.

\begin{singlespace}
\bibliographystyle{elsarticle-num-names}
\bibliography{pohmm}
\end{singlespace}

\clearpage{}

\appendix

\section{Summary of POHMM parameters and variables}
\begin{center}
\begin{table}[h]
\caption{Summary of POHMM parameters and variables.\label{tab:Summary-of-POHMM-variables-parameters}}
\begin{singlespace}
\centering{}\renewcommand{\arraystretch}{1.0}%
\begin{tabular}{|>{\centering}m{0.15\columnwidth}|>{\raggedright}m{0.75\columnwidth}|}
\hline 
\textbf{\scriptsize{}Parameter} & \textbf{\scriptsize{}Description}\tabularnewline
\hline 
{\scriptsize{}$\psi$,$\omega$} & {\scriptsize{}Event types}\tabularnewline
\hline 
{\scriptsize{}$i$, $j$} & {\scriptsize{}Hidden states}\tabularnewline
\hline 
{\scriptsize{}${\bf x}_{1}^{N}$ } & {\scriptsize{}Observation sequence; ${\bf x}_{n}$ is the feature
vector observed at time $t_{n}$}\tabularnewline
\hline 
{\scriptsize{}$\Omega_{1}^{N}$ } & {\scriptsize{}Event type sequence; $\Omega_{n}$ is the event type
observed at time $t_{n}$}\tabularnewline
\hline 
{\scriptsize{}$z_{1}^{N}$ } & {\scriptsize{}Sequence of hidden (unobserved) states; $z_{n}$ is
the hidden state at time $t_{n}$}\tabularnewline
\hline 
{\scriptsize{}$M$ } & {\scriptsize{}Number of hidden states}\tabularnewline
\hline 
{\scriptsize{}$m$ } & {\scriptsize{}Number of unique event types in $\Omega_{1}^{N}$}\tabularnewline
\hline 
{\scriptsize{}$a\left[i,j|\psi,\omega\right]$ } & {\scriptsize{}Probability of transitioning from state $i$ to $j$,
given event types $\psi$ while in state $i$ and $\omega$ in state
$j$}\tabularnewline
\hline 
{\scriptsize{}$\pi\left[j|\omega\right]$ } & {\scriptsize{}Probability of state $j$ at time $t_{1}$, given event
type $\omega$}\tabularnewline
\hline 
{\scriptsize{}$\Pi\left[j|\omega\right]$ } & {\scriptsize{}Stationary probability of state $j$, given event type
$\omega$}\tabularnewline
\hline 
{\scriptsize{}${\bf b}\left[j|\omega\right]$ } & {\scriptsize{}Emission distribution parameters of state $j$, given
event type $\omega$}\tabularnewline
\hline 
{\scriptsize{}$\gamma_{n}\left[j|\omega\right]$ } & {\scriptsize{}Probability of state $j$ at time $t_{n}$, given event
type $\omega$}\tabularnewline
\hline 
{\scriptsize{}$\xi_{n}\left[i,j|\psi,\omega\right]$ } & {\scriptsize{}Probability of transitioning from state $i$ at time
$t_{n}$ to state $j$ at time $t_{n+1}$, given event types $\psi$
and $\omega$ at times $t_{n}$ and $t_{n+1}$, respectively}\tabularnewline
\hline 
\end{tabular}
\end{singlespace}
\end{table}
\par\end{center}

\section{Proof of convergence \label{sec:Proof-of-convergence}}

The proof of convergence follows that of \citet{levinson1983introduction}
which is based on \citet{baum1970maximization}. Only the parts relevant
to the POHMM are described. Let $Q\left(\theta,\dot{\theta}\right)$
be Baum's auxiliary function, 
\begin{equation}
Q\left(\theta,\dot{\theta}\right)=\sum_{z_{1}^{N}\in Z}\ln u_{z_{1}^{N}}\ln v_{z_{1}^{N}}\label{eq:baums-aux}
\end{equation}
where $u_{z_{1}^{N}}=P\left({\bf x}_{1}^{N},z_{1}^{N}|\Omega_{1}^{N},\theta\right)$,
$v_{Z}=P\left({\bf x}_{1}^{N},z_{1}^{N}|\Omega_{1}^{N},\dot{\theta}\right)$,
and $Z$ is the set of all state sequences of length $N$. By Theorem
2.1 in Baum's proof \citep{baum1970maximization}, maximizing $Q\left(\theta,\dot{\theta}\right)$
leads to increased likelihood, unless at a critical point, in which
case there is no change.

Using the POHMM parameters $\dot{\theta}$, $\ln v_{z_{1}^{N}}$ can
be written as 
\begin{eqnarray}
\ln v_{z_{1}^{N}} & = & \ln P\left(z_{1}^{N},{\bf x}_{1}^{N}|\Omega_{1}^{N},\dot{\theta}\right)\nonumber \\
 & = & \ln\dot{\pi}\left[z_{1}|\Omega_{1}\right]+\sum_{n=1}^{N-1}\ln\dot{a}\left[z_{n},z_{n+1}|\Omega_{n},\Omega_{n+1}\right]+\label{eq:baum0}\\
 &  & \sum_{n=1}^{N}\ln f\left({\bf x}_{n};\dot{{\bf b}}\left[z_{n}|\Omega_{n}\right]\right)
\end{eqnarray}
and similarly for $\ln u_{z_{1}^{N}}$. Then, 
\begin{eqnarray}
Q\left(\theta,\dot{\theta}\right) & = & \sum_{z_{1}^{N}\in Z}\bigg\{\ln\dot{\pi}\left[z_{1}|\Omega_{1}\right]+\sum_{n=1}^{N-1}\ln\dot{a}\left[z_{n},z_{n+1}|\Omega_{n},\Omega_{n+1}\right]\nonumber \\
 &  & +\sum_{n=1}^{N}\ln f\left({\bf x}_{n};\dot{{\bf b}}\left[z_{n}|\Omega_{n}\right]\right)\bigg\} P\left(z_{1}^{N}|{\bf x}_{1}^{N},\Omega_{1}^{N},\theta\right)\label{eq:baum1}
\end{eqnarray}
and regrouping terms,

\begin{eqnarray}
Q\left(\theta,\dot{\theta}\right) & = & \sum_{z_{1}\in Z}\ln\dot{\pi}\left[z_{1}|\Omega_{1}\right]P\left(z_{1}|{\bf x}_{1}^{N},\Omega_{1}^{N},\theta\right)\nonumber \\
 &  & +\sum_{z_{n}^{n+1}\in Z}\sum_{n=1}^{N-1}\ln\dot{a}\left[z_{n},z_{n+1}|\Omega_{n},\Omega_{n+1}\right]P\left(z_{n}^{n+1}|{\bf x}_{1}^{N},\Omega_{1}^{N},\theta\right)\nonumber \\
 &  & +\sum_{z_{n}\in Z}\sum_{n=1}^{N}\ln f\left({\bf x}_{n};\dot{{\bf b}}\left[z_{n}|\Omega_{n}\right]\right)P\left(z_{n}|{\bf x}_{1}^{N},\Omega_{1}^{N},\theta\right)\;.\label{eq:baum2}
\end{eqnarray}
Finally, substituting in the model parameters and variables gives,

\begin{eqnarray}
Q\left(\theta,\dot{\theta}\right) & = & \sum_{j=1}^{M}\gamma_{1}\left[j|\Omega_{1}\right]\ln\dot{\pi}\left[j|\Omega_{1}\right]\nonumber \\
 &  & +\sum_{j=1}^{M}\sum_{i=1}^{M}\sum_{n=1}^{N-1}\xi_{n}\left[i,j|\Omega_{n},\Omega_{n+1}\right]\ln\dot{a}\left[i,j|\Omega_{n}|\Omega_{n+1}\right]\nonumber \\
 &  & +\sum_{j=1}^{M}\sum_{n=1}^{N}\gamma_{n}\left[j|\Omega_{n}\right]\ln f\left({\bf x}_{n};\dot{{\bf b}}\left[z_{n}|\Omega_{n}\right]\right)\label{eq:baum3}
\end{eqnarray}

The POHMM re-estimation formulae (Equations \ref{eq:update-startproba-pohmm},
\ref{eq:update-transmat-pohmm}, \ref{eq:update-emission-pohmm})
follow directly from the optimization of each term in Equation \ref{eq:baum3}.
Even when parameter smoothing is used, convergence is still guaranteed.
This is due to the diminishing effect of the marginal for each parameter,
$\lim_{N\to\infty}\tilde{\theta}=\theta$, where $\tilde{\theta}$
are the smoothed parameters.
\end{document}